\documentclass[letterpaper, twocolumn, 10pt]{article}

\usepackage[T1,T2A]{fontenc}
\usepackage[utf8]{inputenc}

\usepackage{adjustbox}
\usepackage{amsfonts}
\usepackage{amsmath}
\usepackage{amssymb}
\usepackage[russian, english]{babel}
\usepackage{blindtext}
\usepackage{booktabs}
\usepackage{caption}
\usepackage{fontawesome}
\usepackage{graphicx}
\usepackage[colorlinks]{hyperref}
\usepackage{listings}
\usepackage{mathtools}
\usepackage{mdwlist}
\usepackage{pgfplots}
\usepackage{relsize}
\usepackage{soul}
\usepackage{subcaption}
\usepackage{tikz}
\usepackage{tikzpeople}
\usepackage{usenix}
\usepackage{url}
\usepackage{xcolor}
\usepackage{tabularx}

\expandafter\def\expandafter\UrlBreaks\expandafter{\UrlBreaks
  \do\a\do\b\do\c\do\d\do\e\do\f\do\g\do\h\do\i\do\j%
  \do\k\do\l\do\m\do\n\do\o\do\p\do\q\do\r\do\s\do\t%
  \do\u\do\v\do\w\do\x\do\y\do\z\do\A\do\B\do\C\do\D%
  \do\E\do\F\do\G\do\H\do\I\do\J\do\K\do\L\do\M\do\N%
  \do\O\do\P\do\Q\do\R\do\S\do\T\do\U\do\V\do\W\do\X%
  \do\Y\do\Z\do\*\do\-\do\~\do\'\do\"\do\-}%

\microtypecontext{spacing=nonfrench}
\pgfplotsset{compat=1.16}

\usetikzlibrary{arrows,arrows.meta,shapes,automata,patterns,petri,positioning,calc,fit,backgrounds,matrix,shadows}

\definecolor{arm-grey}{HTML}{7d868c}
\definecolor{arm-blue}{HTML}{0091bd}
\definecolor{arm-green}{HTML}{95d600}
\definecolor{arm-orange}{HTML}{ff6b00}

\newcommand{\armlabel}{\({}^\dagger\)}
\newcommand{\byte}{B}
\newcommand{\circlenumber}[1]{\raisebox{0.9pt}{\textcircled{\raisebox{0.5pt}{\smaller[2]{}#1}}}}
\newcommand{\circletwonumbers}[2]{\raisebox{0.9pt}{\textcircled{\raisebox{0.5pt}{\smaller[3]{}#1\!\({}_{#2}\)}}}}
\newcommand{\coreiseven}{\emph{Core~i7}}
\newcommand{\deffont}[1]{\textbf{#1}}
\newcommand{\ectwo}{\emph{EC2}}

\newcommand{\gibi}{Gi}
\newcommand{\giga}{G}
\newcommand{\IClabel}{\({}^\mathparagraph\)}

\newcommand{\iotexlabel}{\({}^\star\)}
\newcommand{\kilo}{K}
\newcommand{\kibi}{Ki}

\newcommand{\mebi}{Mi}
\newcommand{\rpifour}{\emph{RPi4}}
\newcommand{\SI}{}
\newcommand{\tera}{T}
\newcommand{\UCSDlabel}{\({}^\ddagger\)}

\newcommand{\spacehack}[1]{\vspace*{-#1}}

\title{Private delegated computations using strong isolation}

\author{
{\rm Mathias Brossard\armlabel, Guilhem Bryant\armlabel, Basma El Gaabouri\armlabel, Xinxin Fan\iotexlabel, Alexandre Ferreira\armlabel}\\
{\rm Edmund Grimley-Evans\armlabel, Christopher Haster\armlabel, Evan Johnson\UCSDlabel, Derek Miller\armlabel, Fan Mo\IClabel}\\
{\rm Dominic P.~Mulligan\armlabel, Nick Spinale\armlabel, Eric van Hensbergen\armlabel, Hugo J.~M.~Vincent\armlabel, Shale Xiong\armlabel}\\
    \\
    \armlabel{}Systems Group, Arm Research
    \quad
    \iotexlabel{}IoTeX.io
    \\
    \UCSDlabel{}University of California, San Diego
    \quad
    \IClabel{}Imperial College London
}%
\date{}

\begin{document}

\maketitle

\begin{abstract}
Sensitive computations are now routinely \emph{delegated} to third-parties.
In response, \emph{Confidential Computing} technologies are being introduced to microprocessors, offering a protected processing environment, which we generically call an \emph{isolate}, providing confidentiality and integrity guarantees to code and data hosted within---even in the face of a privileged attacker.
Isolates, with an attestation protocol, permit remote third-parties to establish a trusted ``beachhead'' containing known code and data on an otherwise untrusted machine.
Yet, the rise of these technologies introduces many new problems, including: how to ease provisioning of computations safely into isolates; how to develop distributed systems spanning multiple classes of isolate; and what to do about the billions of ``legacy'' devices without support for Confidential Computing?

Tackling the problems above, we introduce \emph{Veracruz}, a framework that eases the design and implementation of complex privacy-preserving, collaborative, delegated computations among a group of mutually mistrusting principals.
Veracruz supports multiple isolation technologies and provides a common programming model and attestation protocol across all of them, smoothing deployment of delegated computations over supported technologies.
We demonstrate Veracruz in operation, on private in-cloud object detection on encrypted video streaming from a video camera.
In addition to supporting hardware-backed isolates---like AWS Nitro Enclaves and Arm\textsuperscript{\textregistered} Confidential Computing Architecture Realms---Veracruz also provides pragmatic ``software isolates'' on Armv8-A devices without hardware Confidential Computing capability, using  the high-assurance \emph{seL4} microkernel and our \emph{IceCap} framework.
\end{abstract}

\section{Introduction}
\label{sect.introduction}

Code and data are now routinely shared with a \emph{delegate} who is better placed, either through economies of scale, or computational capacity, to host a computation.
While Cloud computing is the obvious exemplar of this trend, other forms of distributed computing---including volunteer Grid Computing, wherein machines lend spare computational capacity to realize some large computation, and Ambient Computing, wherein computations are \emph{mobile} and hop from device-to-device as computational contexts change---also see computations freely delegated to third parties.

At present, in the absence of the widespread deployment of Advanced Cryptography~\cite{10.1145/3243734.3268995}, delegating computation to a third party inexorably means entering into a trust relationship with the delegate, and for some especially sensitive computations this may be simply unacceptable.
Yet, even for less sensitive delegated computations, there is still an interest in limiting the scope of this trust relationship.
In the Cloud context, though established hosts may be reputable, technical means may be desired to shield computations from prying or interference which may originate from many sources, not only from the hosting company themselves: malefactors may exploit hypervisor bugs to spy on co-tenants, for example.
Cloud hosts also increasingly see an interest in \emph{deniable hosting}, wherein technical measures ensure that a customer's computations simply cannot be interfered with, or spied upon, by the hosts themselves---even in the face of legal compulsion.
For Ambient and volunteer Grid Computing, these concerns also manifest: nodes must be assumed hostile and assumed to be trying to \emph{undermine} a computation, either through malice or as a consequence of bugs or glitches.
As a result, volunteer Grid Computing deployments may schedule computations on multiple nodes and check for consistency~\cite{1630798}.

In response, novel \emph{Confidential Computing} technologies are being added to microprocessor architectures and cloud infrastructure, providing protected computing environments---variously called Secure Enclaves, Realms, Trusted Execution Environments, and which we generically call \emph{isolates}---that provide strong confidentiality and integrity guarantees to code and data hosted within, even in the face of a privileged attacker.
Isolates are also typically paired with an \emph{attestation} protocol, allowing a third-party to deduce, with high confidence, that a remote isolate is authentic and configured in a particular way.
Taken together, one may establish a protected ``beachhead'' on an untrusted third-party's machine---exactly what is needed to protect delegated computations.

Isolates offer a range of benefits for system designers, namely allowing programmers to design arbitrarily complex privacy-preserving distributed systems using standard tools and programming idioms that run at close to native speed.
Moreover, compared to cryptographic alternatives, Confidential Computing technology is available for use and deployment in real systems \emph{today}.
Yet, the emergence of Confidential Computing technology poses some interesting problems.

First, note that Confidential Computing technologies simply provide an empty, albeit secure, isolate.
Associated questions like how computations are securely provisioned into an isolate, how to make this process straightforward and foolproof, and how systems are designed and built around isolates as a new kind of primitive, are left unanswered.
Moreover, for some types of distributed system---such as Grid and Ambient computing systems, previously discussed---it is feasible that \emph{different} types of isolate will be used within a single larger system.
Here, bridging differences in attestation protocol and programming model will be key, as will be easing deployment and scheduling of computations hosted within isolates.

For this reason, we introduce our main research contribution: \emph{Veracruz}, a framework that abstracts over isolates and their associated attestation processes.
Veracruz supports multiple different isolation technologies, including hardware-backed isolates like AWS Nitro Enclaves and Arm Confidential Computing Architecture Realms on a private branch.
Adding support for more is straightforward.
Veracruz provides a uniform programming model across different supported isolates---using WebAssembly (Wasm, henceforth)\cite{webassembly}---and a generalized form of attestation, providing a ``write once, isolate anywhere'' style of development: programs can be protected using \emph{any} supported isolation technology without recompilation.
Veracruz is discussed in \S\ref{sect.veracruz}.

Veracruz captures a particularly general form of interaction between mutually mistrusting parties.
As a result, Veracruz can be specialized in a straightforward manner to obtain an array of delegated, privacy-preserving computations of interest.
In support of this claim we provide a description of how Veracruz can be used for secure ML model aggregation, and an industrial case-study built around AWS Nitro Enclaves, demonstrating an end-to-end encrypted video decoding and object-detection flow, using a deep learning framework processing video obtained from an IoT camera.
These case-studies, and further benchmarking, are discussed in \S\ref{sect.evaluation}.

In \S\ref{sect.hardware-backed.confidential.computing} we argue that Confidential Computing technology is likely to be widely deployed within industry, despite well-known flaws in particular implementations.
Yet, \emph{billions} of existing devices have already been shipped without any explicit support for Confidential Computing, and these devices will continue to be used for years, if not decades, to come.
Is there some \emph{pragmatic} isolation mechanism that we could use on ``legacy'' devices which, while falling short of the confidentiality and integrity guarantees offered by hardware-backed Confidential Computing mechanisms, can yet provide believable isolation for workloads?
Rising to this challenge, we introduce our second research contribution: \emph{IceCap}, a pragmatic ``software isolate'' for Armv8-A devices without explicit support for Confidential Computing.
IceCap uses the high-assurance \emph{seL4} microkernel to provide strong confidentiality and integrity guarantees for VMs, with little overhead.

IceCap is supported by Veracruz, and taken together, one may design and deploy delegated computations across hardware- and software-isolates on next-generation and legacy hardware, alike.
We introduce IceCap in \S\ref{sect.icecap}, as a stepping stone to the introduction of Veracruz.

\section{Hardware-backed Confidential Computing}
\label{sect.hardware-backed.confidential.computing}

In addition to the already widely-deployed \emph{Arm TrustZone\textsuperscript{\textregistered}}~\cite{arm:trustzone:2020} and \emph{Intel Software Guard Extensions} (SGX)~\cite{costan:intel:2016}, an emerging group of novel \emph{Confidential Computing} technologies are being added to microprocessor architectures and cloud infrastructures, including \emph{AMD Secure Encrypted Virtualization} (SEV)~\cite{kaplan:amd:2016}, \emph{Arm Confidential Computing Architecture} (CCA)~\cite{arm-cca}, \emph{AWS Nitro Enclaves}~\cite{AWSNitroAttestation}, and \emph{Intel Trust Domain Extensions} (TDX)~\cite{intel:tdx}.
All introduce a hardware-backed protected execution environment, which we call an \deffont{isolate}, providing strong \deffont{confidentiality} (the content of the isolate remains opaque to external observers) and \deffont{integrity} (the content of the isolate remains protected from interference by external observers) guarantees to code and data hosted within.
These guarantees apply even in the face of a strong adversary, with any operating system or, in most cases even a hypervisor, outside of the isolate assumed hostile.
Memory encryption may also be provided as a standard feature to protect against a class of physical attack.
Isolates are often associated with an \deffont{attestation protocol}---e.g., \emph{EPID} for Intel SGX~\cite{5591478, DBLP:journals/tdsc/BrickellL12} and \emph{AWS Nitro Attestation} for AWS Nitro Enclaves~\cite{AWSNitroAttestation}.
These permit a third party to garner strong, cryptographic evidence of the authenticity and configuration of a remote isolate.

Some isolate implementations have unfortunately fallen short of their promised confidentiality and integrity guarantees.
A substantial body of academic work, demonstrating that side-channel (see e.g.~\cite{sgx-step, 7163052, 203868, 10.1145/3243734.3243822, DBLP:journals/tches/HuoMWHZZL20, DBLP:conf/ches/MoghimiIE17, 206170, DBLP:journals/tches/DallMEGHMY18, 8806740, DBLP:conf/infocom/00170SLH18}) and fault injection attacks~\cite{Murdock2019plundervolt, 203864, 263816} can be used to exfiltrate secrets from isolates, now exists, and a perception---at least in the academic community and technical press---appears to be forming that isolates are fundamentally broken and any consequent research project that builds upon them need necessarily justify that decision.
We argue that this emerging perception is an instance of \emph{the perfect being the enemy of the good}.

First, we expect that many identified flaws will be gradually ironed out over time, either in point-fixes, iterated designs, or by the adoption of software models that avoid known vulnerabilities.
For hardware, we have already seen some flaws fixed using microcode updates and other point-fixes by affected manufacturers (e.g,~\cite{sgx-platypus-microcode}).
For software, research into methods designed to avoid known classes of side-channels is emerging, through implementation techniques such as constant-time algorithms, and dedicated type-systems such as FaCT~\cite{FaCT} and CT-Wasm~\cite{ct-wasm}.
These may prove to be useful in implementing systems with isolates, and we summarize our own ongoing experimentation with these approaches in \S\ref{sect.closing.remarks}.

Second, we expect that industrial adoption of isolates will be widespread, and arguably this is already in evidence with the formation of consortia such as the LF's \emph{Confidential Computing Consortium}~\cite{ccc}, and an emerging ecosystem of industrial users and startups.
Researching systems that use isolates, and ease their deployment, is therefore not only justifiable, but very useful.
Here, industrial users pragmatically evaluate isolate-based systems in comparison with the \emph{status quo}, where delegated computations are---by and large---left completely unprotected, and we argue that it is this standard which should be applied when evaluating systems built around isolates, not comparison with side-channel free cryptography which is still impractical in an industrial context.
In this light, forcing malefactors to resort to side-channel and fault injection attacks---many of which are impractical, or can be defended against using others means---to exfiltrate data from an isolate is a welcome, albeit incremental, improvement in the privacy-guarantees that real systems can offer users.

\section{IceCap}
\label{sect.icecap}

\emph{IceCap} is a hypervisor with a minimal trusted computing base (TCB, henceforth) built around the formally verified \emph{seL4} microkernel.
IceCap provides a pragmatic and flexible \emph{software isolate} for many existing Armv8-A devices.
The IceCap hypervisor relegates the untrusted operator to a domain of limited privilege called the host.
This domain consists of a distinguished virtual machine---housing a rich operating system such as Linux---and a minimal accompanying virtual machine monitor.
The host domain manages the device's CPU and memory resources, and drives device peripherals which the TCB does not depend on.
This includes opaque memory and CPU resources for confidential virtual machines---or isolates.
However, the host does not have the right to access the resources of isolates---while scheduling and memory management \emph{policy} is controlled by the host, \emph{mechanism} is the responsibility of more trustworthy components.

IceCap's TCB includes the seL4 microkernel and compartmentalized, privileged seL4-native services running in EL0.
These co-operate defensively with the host to expose isolate lifecycle, scheduling, and memory management mechanisms.

At system initialization, the hypervisor extends from the device's root of trust via a device-specific measured boot process and then passes control to the untrusted host domain.
A remote party coordinates with the host to spawn a new isolate by first sending a declarative specification of the isolate's initial state to IceCap's \emph{trusted spawning service}, via the host, which then carves-out the requested memory and CPU resources from resources which are inaccessible to the host.
A process on the host, called the \emph{shadow virtual machine monitor}, provides untrusted paravirtualized device backends to isolates, and also acts as a \emph{token} representing the isolate in the host's scheduler, to enable the host operating system to manage isolate scheduling policy with minimal modification.

To support attestation of isolates, IceCap would use a platform-specific measured boot to prove its own identity and then attest that of an isolate to a remote challenger.
This is not yet implemented, with IceCap attestation being stubbed to support Veracruz, but straightforward to do so.

seL4 is accompanied by security and functional correctness proofs, checked in \emph{Isabelle/HOL}~\cite{DBLP:conf/itp/SewellWGMAK11, DBLP:conf/cpp/MurrayMBGK12, DBLP:conf/sp/MurrayMBGBSLGK13}, providing assurance that IceCap correctly protects isolates from software attacks.
By using seL4, IceCap will also benefit from ongoing research into the elimination of certain classes of timing channels~\cite{sel4-timing-protection}.
The trusted seL4 userspace components of IceCap are not yet verified, though they are compartmentalized and initialized using \emph{CapDL}~\cite{capdl}, which has a formal semantics known to be amenable to verification~\cite{capdl-semantics} from previous work.
Using the high-level seL4 API, these components are also implemented at a high level of abstraction in Rust, making auditing easier and eliminating the need to subvert the Rust compiler's memory safety checks---even for components which interact with hardware address translation structures.
The IceCap TCB is small and limited in scope---about $40,000$ lines of code.
Virtual machine monitors are moved to the trust domains of the virtual machines they supervise, thereby eliminating emulation code from the TCB.
Towards that end, cross-domain fault handling is replaced with higher-level message passing via seL4 IPC.

Isolates are also protected with the System MMU (SMMU) from attacks originating from peripherals under the host's control.
IceCap is designed to seamlessly take advantage of additional hardware security features based on, or aligned with, address translation-based access controls---Arm TrustZone~\cite{arm:trustzone:2020}, for example.
TrustZone firmware typically uses the \texttt{NS} state bit to implement a coarse context switch, logically partitioning execution on the application processor into two \emph{worlds}.
IceCap could use this to run isolates out of secure-world memory resources, protected by platform-specific mechanisms which may mitigate certain classes of physical attack.


Under IceCap, isolate and host incur a minimal performance overhead compared to host and guests under \emph{KVM}~\cite{kvm}.
We use \emph{Firecracker}~\cite{firecracker}---an open-source VMM for KVM from AWS---as a point of comparison, due to its minimalism for the sake of performance, and preference for paravirtualization over emulation.
Compute-bound workloads in IceCap isolates incur a ${\sim}2.2$\% overhead compared to native Linux processes and a ${\sim}1.8$\% overhead compared to Firecracker guests due to context switches through the TCB on timer ticks (see Table~\ref{fig.icecap.overheads}).
The virtual network bandwidth between the host and an isolate represents how data flows through IceCap in bulk.
However, at the time of writing, untrusted network device emulation differs from Firecracker's trusted network device emulation in ways that hinder a satisfying comparison, and with this in mind, we note guest-to-host incurs a ${\sim}9.9$\% bandwidth overhead, whereas host-to-guest outperforms Firecracker by a small margin.
As IceCap's implementation matures, we expect virtual network bandwidth overhead to settle between these two points.

\begin{table}[t!]
\centering
\small
\begin{tabular}{rl@{\quad}l} \toprule
& \multicolumn{2}{c}{Events per second (via \texttt{sysbench})} \\ \cmidrule(l){2-3}
& Host & Guest \\ \midrule
\emph{Firecracker} & 586.18 & 582.65 (-0.60\%) \\
\emph{IceCap} & 583.68 (-0.43\%) & 572.28 (-2.18\%) \\ \midrule
& \multicolumn{2}{c}{Bandwidth (Gbits/sec)} \\ \cmidrule(l){2-3}
& Guest $\rightarrow$ Host & Host $\rightarrow$ Guest \\ \midrule
\emph{Firecracker} & 3.42 & 3.14 \\
\emph{IceCap} & 3.08 (-9.9\%) & 3.18 (+1.3\%) \\ \bottomrule
\end{tabular}
\caption{Overheads for IceCap compute-bound workloads (top) and virtual network performance (bottom)}
\label{fig.icecap.overheads}
\end{table}

The great performance of seL4 IPC~\cite{sel4-ipc} helps reduce IceCap's performance overhead, and this is further helped by minimizing VM exits using aggressive paravirtualization: VMMs for both host and guest do not even map any of their VMs' memory into their own address spaces, and their only runtime responsibility is emulating the interrupt controller, with their VMs employing interrupt mitigation to even avoid that.

Next, we introduce a framework for designing and deploying privacy-preserving delegated computations across various different isolation technologies---IceCap included.




\section{Veracruz}
\label{sect.veracruz}


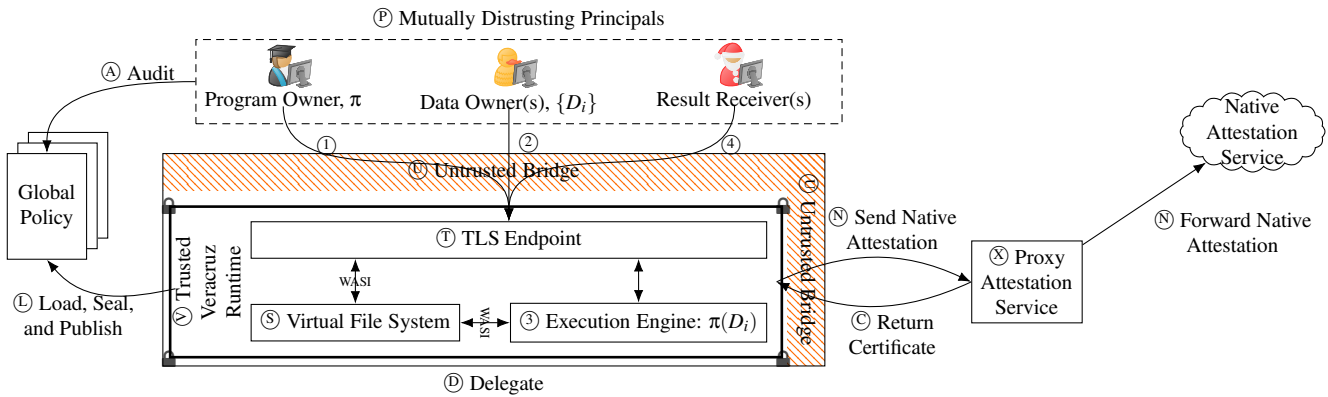
\begin{figure*}[t]
\centering
\begin{tikzpicture}[
    doc/.style={draw, minimum height=4em, minimum width=3em,
                fill=white,
                double copy shadow={shadow xshift=4pt,
                             shadow yshift=4pt, fill=white, draw}},
    singlearrow/.style={-Latex},
    doublearrow/.style={Latex-Latex},
    font=\footnotesize,
]
\node[graduate,female,stripes=arm-blue,monitor] (program) {Program Owner, \( \pi \)};
\node[duck,hair=skin,monitor] (data) at    ($(program) + (3,0)$) {Data Owner(s), \(\{ D_i \}\)};
\node[santa,monitor] (result)  at ($(data) + (3,0)$) {Result Receiver(s)};
\node (resultright) at ($(result.base east) + (1,0)$) {};
\node[fit=(program)(data)(result)(program.text)(data.text)(result.text)(resultright),draw,densely dashed,label={[name=clientlabel]above:\circlenumber{P} Mutually Distrusting Principals}] (clients) {};

\node[pattern=north west lines, pattern color=arm-orange,text=black,minimum width=250pt] (bridge) at ($(data) + (-0.2,-1.4)$) {\circlenumber{U} Untrusted Bridge};
\node[draw,minimum width=195pt,fill=white] (tls)  at ($(bridge) + (0.2,-0.9)$) {\circlenumber{T} TLS Endpoint};
\node[draw,minimum width=40pt,fill=white,align=center,anchor=west] (vfs)  at ($(tls.west) + (0,-1.1)$) {\circlenumber{S} Virtual File System};
\node[draw,minimum width=40pt,fill=white,align=center,anchor=east] (vm)  at ($(tls.east) + (0,-1.1)$) {\circlenumber{3} Execution Engine: $\pi(D_i)$};
\node[fit=(vfs)(tls)(vm),label={[name=applabel,rotate=90,anchor=south,align=center,inner sep=0pt]left:\circlenumber{V} Trusted\\Veracruz\\Runtime}] (appinner) {};
\node[line width=1pt,fit=(appinner)(applabel),draw,inner sep=2pt] (app) {};
\node[pattern=north west lines, pattern color=arm-orange,text=black,rotate=-90,minimum width=80pt,anchor=north west] (bridge-right) at (bridge.north east) {\circlenumber{U} Untrusted Bridge};
\node[draw,fit=(app)(applabel)(bridge)(bridge-right),inner sep=0pt, label={[name=targetlabel]below:\circlenumber{D} Delegate}] (target) {};

\node[opacity=0.7] at (app.north east) {\faLock};
\node[opacity=0.7] at (app.south east) {\faLock};
\node[opacity=0.7] at (app.north west) {\faLock};
\node[opacity=0.7] at (app.south west) {\faLock};

\node[doc,align=center,anchor=north east] (policy) at ($(target.north west) + (-1,0)$) {Global\\Policy};

\node[draw,align=center,fill=white,anchor=west] (proxy) at ($(app.east) + (2.5,0)$) {\circlenumber{X} Proxy\\Attestation\\Service};
\node[cloud, draw, align=center, cloud puffs=20,cloud puff arc=110, aspect=2, inner sep=-1mm] (native) at ($(proxy) + (3.0,2)$) {Native\\Attestation\\Service};

\draw[singlearrow] (program.base) to[out=-90,in=90] node[pos=0.2,right] {\circlenumber{1}} (tls.north);
\draw[singlearrow] (data.base) to[out=-90,in=90] node[pos=0.3,right] {\circlenumber{2}} (tls.north);
\draw[singlearrow] (result.base) to[out=-90,in=90] node[pos=0.2,right] {\circlenumber{4}} (tls.north);
\draw[singlearrow] (appinner.east) to[out=25,in=155] node[pos=0.6,above,anchor=south,align=center]{\circlenumber{N} Send Native\\Attestation} (proxy.west);
\draw[singlearrow] (proxy) to node[midway,below,align=center, anchor=north west] {\circlenumber{N} Forward Native\\Attestation} (native);
\draw[singlearrow] (proxy.west) to[out=-155,in=-25] node[pos=0.4,below,anchor=north,align=center]{\circlenumber{C} Return\\Certificate} (appinner.east);

\draw[doublearrow] (vfs |- tls.south) to node[pos=0.5,inner sep=0pt,align=center] {\tiny{}WASI} (vfs);
\draw[doublearrow] (vm |- tls.south) to (vm);
\draw[doublearrow] (vfs) to  node[pos=0.5,inner sep=0pt,align=center,rotate=-90] {\tiny{}WASI} (vm);

\draw[singlearrow] (clients) to[in=90,out=180] node[pos=0.1,inner sep=0pt,anchor=south east,align=center] {\circlenumber{A} Audit} (policy);
\draw[singlearrow] (applabel) to[in=-50,out=150] node[pos=0.3,inner sep=0pt,anchor=north east,align=center] {\circlenumber{L} Load, Seal,\\and Publish} (policy.south);

\end{tikzpicture}%
%
\spacehack{5pt}
\caption[Veracruz component diagram]%
{An overview of an abstract Veracruz computation, showing principals and their roles, major system components, and a suggestive depiction of data-flow.
Isolates, such as those hosting the Veracruz runtime, are marked with boxes with padlocks
}
\label{fig:components}
\label{fig:veracruz_overview}
\end{figure*}

Throughout this section we make reference to the system components presented in the schematic in Fig.~\ref{fig:veracruz_overview}.

Veracruz is a \emph{framework} which may be specialized to obtain a particular privacy-preserving, collaborative computation of interest.
A Veracruz computation involves an arbitrary number of \deffont{data owners}, trying to collaborate with a single \deffont{program owner}.
The framework places no limits on the number of data owners, but a particular computation obtained by specializing Veracruz will always spell out a precise number of participants.
We use $\pi$ to denote the program of the program owner, and use $D_i$ for $1 \leq i \leq N$ to denote the data sets of the various data owners in an arbitrary Veracruz computation.

Collectively, the goal of the various principals, \circlenumber{P}, is straightforward: they wish to compute the value $\pi(D_1, \ldots, D_N)$, that is, the value of the program $\pi$ applied to the $N$ inputs of the various data owners.
To do this, they may choose to make use of a third party machine to power the computation, \circlenumber{D}.
We refer to the owner of this machine as the \deffont{delegate}, and this machine is assumed capable of launching an isolate of a type that Veracruz supports, loaded with the Veracruz \deffont{trusted runtime}, \circlenumber{V}.
This runtime acts as a ``neutral ground'' within which a computation takes place, and provides strong \deffont{sandboxing} guarantees to the delegate, who is loading untrusted code in the form of $\pi$, onto their machine.
The runtime is open-source, and auditable by principals, assuming bit-for-bit reproducible builds.

Each principal in a Veracruz computation has a mixture of \deffont{roles}, consisting of some combination of \deffont{data provider}, \deffont{program provider}, \deffont{delegate}, and \deffont{result receiver}.
While the first three have been implicitly introduced, the latter role refers to principals who will receive the result of the computation.
The identification details of each principal, in the form of cryptographic certificates (or an IP address for the delegate), and their mixture of roles, is captured in a public \deffont{global policy} configuration file, \circlenumber{L}, which parameterizes each computation, and which also contains other important bits of metadata.
Only one principal may be delegate or program provider.

The global policy captures the \deffont{topology} of a computation, specifying where information may flow from, and to whom, in a computation, while varying the program $\pi$ varies precisely \deffont{what} is being computed.
By varying the two, Veracruz can capture a general pattern of interaction shared by many delegated computations, and one could, for example, effect a varied palette of computations of interest, including:

\noindent
\emph{Moving heavy computations safely off a computationally-weak device to an untrusted edge device or server.}
The computationally-weak device is both data provider and result receiver, the untrusted edge device or server is delegate, and the computationally-weak device or its owner is the program provider, providing the computation to be performed.

\noindent
\emph{Privacy-preserving machine learning between a pair of mutually distrusting parties with private datasets, but where learnt models are made available to both participants.}
Both principals are data providers, contributing their datasets provided in some common format, and also act as result receivers for the learnt model.
Arbitrarily one acts as program provider, providing the implementation of the machine learning algorithm of interest.
A third-party, e.g.,~a Cloud host, acts as delegate.

\noindent
\emph{A DRM mechanism wherein novel IP (e.g.,~computer vision algorithms) are licensed out on a ``per use'' basis, and where the IP is never exposed to customers.}
The IP owner is program provider, and the licensee is both data provider and result receiver, providing the inputs to, and receiving the output from, the private IP.
The IP owner themselves may act as delegate, or this can be contracted out to a third-party.
With this, the IP owner never observes the input or output of the computation, and the licensee never observes the IP.

\noindent
\emph{The implementation of privacy-preserving auctions.}
An auction service acts as program provider, implementing a sealed-bid auction, and also acts as delegate.
Bidders are data providers, submitting sealed bids.
All principals are also result receivers, receiving notice of the auction winner and the price to be paid, which is public.
Neither bidder nor auction service ever learn the details of any bids, other than their own and the winning bid.

In addition, it is easy to see how more complex distributed systems can be built \emph{around} Veracruz.
For example, a volunteer Grid computing framework where confidentiality is not paramount, but computational integrity is; an Ambient computing runtime for mobile computations across a range of devices; a privacy-preserving MapReduce~\cite{mapreduce} or Function-as-a-Service (FaaS, henceforth) style framework.
Here, computational nodes act as an independent delegate for some aspect of the wider computation, and \emph{different} isolation technologies may also be used in a single computation, either due to availability for Grid or Ambient computing, or due to scheduling of sensitive sub-computations onto stronger isolation mechanisms for MapReduce.

In the most general case, each principal in a Veracruz computation is mutually mistrusting, and does not wish to \deffont{declassify}---or intentionally reveal---their data: data providers do not wish to divulge their input datasets and the program provider does not wish to divulge their program.
Nevertheless, as the examples enumerated above indicate, for some computations declassification \emph{can} be useful, for example as inducement to other principals to enroll in the computation, a ``nothing up my sleeve'' demonstration.
Referring back to the privacy-preserving machine learning use-case, above, the program provider may \emph{intentionally} declassify their program for auditing---before other principals agree to participate---as a demonstration that the program implements the correct algorithm, and will not (un)intentionally leak secrets.
Similarly, for a Grid computing project, revealing details of the computation, as an enticement to users to donate their spare computational capacity, may be beneficial.

Declassification can also occur as a side effect of the computation itself, for example when the result of a computation---which can reveal significant amounts of information about its inputs, depending on $\pi$---is shared with an untrusted principal.
Principals must evaluate the global policy carefully, before enrolling, to understand where results will flow to, and what they may say about any secrets.
Though Veracruz can be used to design privacy-preserving distributed computations, not \emph{every} computation is necessarily privacy-preserving.

Once the delegate has spawned an isolate with the Veracruz runtime loaded, the program and data owners establish a TLS connection, using a modified TLS handshake, with the isolate \circlenumber{T}, as will be described later in \S\ref{subsect.veracruz.attestation}.
This handshake assures the principals that the isolate is, in fact, executing the Veracruz runtime specified in the global policy, and that the isolate is the other end of their TLS connection.
Once this TLS channel is established, the program and data providers use it to \deffont{provision} their respective secrets directly into the isolate, \circlenumber{1} and \circlenumber{2}.
This makes use of an untrusted bridge, \circlenumber{U}, on the delegate's machine but outside of the isolate, to forward encrypted TLS data received into the isolate itself.
To the delegate, communication via this bridge is immutable and opaque---except for sizing and timing information that TLS leaks---unless they can subvert TLS.
Note that TLS configuration options, including permitted ciphersuites, and the \texttt{SHA-256} hash of the program $\pi$, are also specified in the global policy.
This latter aspect ensures that when a program, $\pi$, is declassified, it can be audited by other principals, and verified to be the same program provisioned into the isolate.

Provisioned secrets are stored as files in a virtual, \deffont{in-memory filesystem} maintained by the Veracruz runtime, \circlenumber{S}.
The contents of this filesystem never leave the isolate, and are destroyed when the isolate is torn down.
The paths of data inputs, $D_i$, are specified in the global policy file, as the program $\pi$ needs to know where its inputs are stored for processing when the computation starts executing.
Similarly, the program $\pi$ is also stored as a file, and will be read from the filesystem itself when loaded for execution by the runtime.

Once everything is in place, a result receiver may request the result of the computation, triggering the Veracruz runtime to load the provisioned program, $\pi$, into the execution engine, \circlenumber{S}, and either compute the result $\pi(D_1, \ldots, D_N)$, terminate with an error code, or diverge.
Assuming a result is computed, it is stored by the program as a file in the filesystem at a path specified by the global policy.
The runtime reads this path, or fails with an error if the program did not write a result there, and makes the result retrievable securely, via TLS, to all result receivers, \circlenumber{4}.
The computation is now complete.

\subsection{Attestation}
\label{subsect.veracruz.attestation}

\begin{figure*}[t]
\centering
\begin{tikzpicture}[
    singlearrow/.style={-Latex},
    doublearrow/.style={Latex-Latex},
    font=\footnotesize,
]

\node[alice] (principals) {\circlenumber{P} Principals};
\node[draw,align=center] (isolate) at ($(principals)+(3.7,-0.2)$) {\circlenumber{V} Isolate\\key pair: \((k_{\texttt{pub}}, k_{\texttt{pri}})\)};
\node[draw,align=center] (proxy) at ($(isolate)+(6.0,0)$) {\circlenumber{X} Proxy Attestation Service\\root key pair: \((r_{\texttt{pub}}, r_{\texttt{pri}})\)};
\node[cloud, draw, align=center, cloud puffs=20,cloud puff arc=110, aspect=2, inner sep=-1mm] (native) at ($(proxy) + (5.5,0)$) {Native\\Attestation\\Service};
\node[] (PERSON-PADDING) at (principals.mid west) {};

\draw[doublearrow,arm-orange] (proxy) to node[midway,above=5pt,align=center] {\circletwonumbers{O}{2} Native attestation} (native);

\draw[singlearrow,arm-blue] (isolate) to[bend left=10] node[midway,above,align=center] {\circletwonumbers{O}{1} Native attestation request on a fresh challenge\\and certificate signing request on key pair \(k\)} (proxy);
\draw[singlearrow,arm-blue] (proxy) to [bend left=10] node[midway,below,align=right] {\circletwonumbers{O}{3} \texttt{X.509} Certificate for the isolate} (isolate);
\draw[singlearrow,arm-blue] (isolate.south west) to[bend left=20] node[pos=0.5,below,align=center] {\circletwonumbers{O}{4} \texttt{X.509} Certificate for the isolate} (PERSON-PADDING.south east);

\draw[singlearrow,arm-grey] (principals.north east) to[bend left=20] node[pos=0.5,above,align=center]{\circlenumber{R}TLS Handshake} (isolate.north west) ;

\end{tikzpicture}%
%
\spacehack{5pt}
\caption{A schematic diagram of the Veracruz attestation service onboarding and challenge protocols}
\label{fig:proxy_attestation}
\end{figure*}
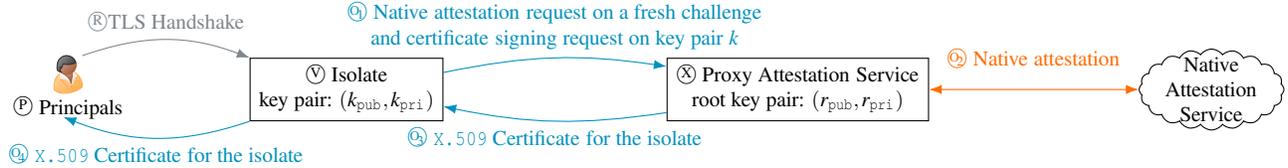

Given Veracruz supports multiple isolation technologies, this poses a series of attestation-related problems: 

\emph{Complex client code}: Software used by principals delegating a computation to Veracruz must support multiple attestation protocols, complicating it.
As Veracruz adds support for more isolation mechanisms---potentially with new attestation protocols---this client code must be updated to interact with the new class of isolate.

\emph{Leaky abstraction}: Veracruz abstracts over isolation technology, allowing principals to easily delegate computations without worrying about the programming or attestation model associated with any one class of isolate.
Forcing clients to switch attestation protocols, depending on the isolation technology, breaks this uniformity.

\emph{Potential side-channel}: For some attestation protocols, each principal in a Veracruz computation must refer attestation evidence to an external attestation service.

\emph{Attestation policy}: principals may wish to disallow computations on delegates with particular isolation technologies.
These policies may stem from security disclosures---vulnerabilities in particular firmware versions, for example---changes in business relationships, or geopolitical trends.
Given our support for heterogeneous isolation technologies, being able to declaratively specify who or what can be trusted becomes desirable.
Existing attestation services do not take policy into account, pushing the burden onto client code---problematic if policy changes, as client code must be updated.

In response, we introduce a \deffont{proxy attestation service} for Veracruz, which must be explicitly trusted by all principals to a computation, with associated server and management software open source, and auditable by anyone.
This service is not protected by an isolate, though in principle could be, and doing so would allow principals to check the authenticity of the proxy attestation service, before trusting it, for example.
Implementing this would be straightforward; for now we assume that the attestation service is trusted, implicitly.

The proxy attestation service first uses an \deffont{onboarding} process to enroll an isolate hosting Veracruz, after which the isolate can act as a TLS server for principals participating in a computation.
We describe these steps, referring to Fig.~\ref{fig:proxy_attestation}.

\paragraph{Onboarding an isolate}
The proxy attestation service maintains a root CA key (a public/private key pair) and a Root CA certificate containing the root CA public key, signed by the root CA private key.
This root CA certificate is included in the global policy file of any computation using that proxy attestation service.
An onboarding protocol is then followed:

\begin{enumerate*}
\item
Upon initialization inside the isolate, the Veracruz runtime \circlenumber{V} generates an asymmetric key pair, along with a \emph{Certificate Signing Request} (or CSR, henceforth)\cite{RFC2986} for that key pair.
\item
The Veracruz runtime performs the platform's \deffont{native attestation flow} \circletwonumbers{O}{1} with the proxy attestation server acting as challenger \circlenumber{X}.
These native attestation flows provide fields for user-defined data, which we fill with a cryptographic hash (\texttt{SHA-256}) of the CSR, which cryptographically binds the CSR to the attestation data, ensuring that they both come from the same isolate.
The Veracruz runtime sends the CSR to the proxy attestation server along with the attestation evidence.
\item
The proxy attestation server \deffont{authenticates} the attestation evidence received via the native attestation flow.
Depending on the particular protocol, this could be as simple as verifying signatures via a known-trusted certificate, or by authenticating the received evidence using an external attestation service.
\item The proxy attestation service computes the hash of the received CSR and compares it against the contents of the user-defined field of the attestation evidence. If it matches, it confirms that the CSR is from the same isolate as the evidence.
\item
The proxy attestation server converts the CSR to an \texttt{X.509} Certificate~\cite{RFC5280} containing a custom extension capturing details about the isolate derived from the attestation process, including a hash of the Veracruz runtime executing inside the isolate (and optionally other information about the platform on which the isolate is executing).
The certificate is signed by the private component of the proxy attestation server's Root CA key.
\item
The proxy attestation server returns the generated certificate to the Veracruz runtime inside the isolate.
\end{enumerate*}

In the typical CA infrastructure, a delegated certificate can be revoked by adding it to a \emph{Certificate Revocation List}, checked by clients before completing a TLS handshake.
While this scheme is possible with our system, we elected to use a different approach, setting the expiry in the isolate's certificate to a relatively short time in the future, so that the proxy attestation service can limit the amount of time a compromised isolate can be used in computations.
The lifetime of isolate certificates can be decided upon via a policy of the proxy attestation service, based upon their appetite for risk.

\paragraph{Augmented TLS handshake}
After an isolate is onboarded, \circletwonumbers{O}{1}, a principal, \circlenumber{R} can attempt to connect to it, using an augmented TLS handshake.
In response to the ``Client Hello'' message sent by the principal, the isolate responds with a ``Server Hello'' message containing the certificate that the isolate received from the proxy attestation server, described above.
The principal then verifies that certificate against the proxy attestation server root CA certificate contained within the global policy.
If it matches, it recognizes that the certificate was indeed generated by the proxy attestation server.
Recall that this certificate contains a custom extension.
Assuming successful verification, the principal then checks the data contained in this extension against the expected values in the global policy.
As currently implemented, the extension contains the hash of the Veracruz runtime, which is also listed in the global policy, and the two are checked by the principal.
If they match, the principal continues the TLS handshake, confident in the fact that it is talking to a Veracruz runtime executing inside of a supported isolation technology.

Note that the proxy attestation service solves the problems with attestation described above.
First, client code is provided with a uniform attestation interface---here, we use Arm's PSA attestation protocol~\cite{Tschofenig_2019}---independent of the underlying isolation technology in use.
Second, none of the principals in the computation need to communicate with any native attestation service.
Thus, the native attestation service knows that software was started in a supported isolate, but it has no knowledge of the identities or even the number of principals.
Finally, the global policy represents the only source of policy enforcement. 
The authors of the global policy can declaratively describe who and what they are willing to trust, with a principal's client software taking this information into account when authenticating or rejecting an attestation token.

Lastly, we note. that our attestation process is specifically designed to accommodate client code running on embedded microcontrollers---e.g.,~Arm Cortex\textsuperscript{\textregistered}-M3 devices---with limited computational capacity, constrained memory and storage (often measured in tens of kilobytes), and which tend to be battery-powered with limited network capacity.
Communication with an attestation service is therefore cost- and power-prohibitive, and using a certificate-based scheme allows constrained devices to authenticate an isolate running Veracruz efficiently.
To validate this, we developed Veracruz client code for microcontrollers, using the Zephyr embedded OS~\cite{zephyr}.
Our client code is \SI{9}{\kilo\byte} on top of the \texttt{mbedtls} stack~\cite{mbedtls}, generally required for secure communication anyway.
Using this, small devices can offload large computations safely to an attested Veracruz instance.

\subsection{Programming model}
\label{subsect.veracruz.programming.model}

Wasm~\cite{wasm-semantics} is designed as a sandboxing mechanism for use in security-critical contexts---namely web browsers---designed to be embeddable within a wider host, has a precise semantics~\cite{DBLP:conf/cpp/Watt18}, is widely supported as a target by a number of high-level programming languages such as Rust and C, and has high-quality interpreters~\cite{wasmi} and JIT execution engines available~\cite{wasmtime}.
We have therefore adopted Wasm as our executable format, supporting both interpretation and JIT execution, with the strategy specified in the global policy.

Veracruz uses Wasm to protect the delegate's machine from the executing program, to provide a uniform programming model, to constrain the behavior of the program, and to act as a portable executable format for programs, abstracting away the underlying instruction set architecture.
Via Wasm, the trusted Veracruz runtime implements a ``two-way isolate'' wherein the runtime is protected from prying and interference from the delegate, and the delegate is protected from malicious program behaviors originating from untrusted code.

To complete a computation, a Wasm program needs some way of reading inputs provided to it by the data provider, and some way of writing outputs to the result receivers.
However, we would like to constrain the behavior of the program as far as possible: a program dumping one of its secret inputs to \texttt{stdout} on the host's machine would break the privacy guarantees that Veracruz aims to provide, for example.
Partly for this reason, we have adopted the WebAssembly System Interface~\cite{wasi} (or Wasi, henceforth) as the programming model for Veracruz.
Intuitively, this can be thought of as ``Posix for Wasm'', providing a system interface for querying Veracruz's in-memory filesystem, generating random bytes, and executing other similar system tasks.
(In this light, the Veracruz runtime can be seen as a simple operating system for Wasm.)
By adopting Wasi, one may also use existing libraries and standard programming idioms when targeting Veracruz.

Wasi uses \deffont{capabilities}, in a similar vein to Capsicum~\cite{DBLP:journals/cacm/WatsonALK12}, and a program may only use functionality which it has been explicitly authorized to use.
The program, $\pi$'s, capabilities are specified in the global policy, and typically extend to reading inputs, writing outputs, and generating random bytes, constraining the program to act as a pure, randomized, function.

\subsection{Ad hoc acceleration}
\label{subsect.veracruz.ad.hoc.acceleration}

Many potential Veracruz applications make use of common, computationally intensive, or security-sensitive routines: cryptography, (de)serialization, and similar.
While these routines could be compiled into Wasm, this may incur a performance penalty compared to optimized native code, and for operations such as cryptography, compilation to Wasm may not preserve security properties such as timing side-channel safety.
Rather, it is beneficial to provide a single, efficient, and correct implementation for common use, rather than routines being compiled into Wasm code haphazardly.

In response, we introduced ``native modules'' providing acceleration for specific tasks which are linked into the Veracruz runtime and invoked from Wasm programs.
In benchmarking one such module---the acceleration of (de)serialization of \texttt{Json} documents from the \texttt{pinecone} binary format---we observe a $35$\% speed-up when (de)serializing a vector of $10,000$ random elements ($238$s native vs.~$375$s Wasm).
Additional optimization will likely further boost performance.

Given the \emph{ad hoc} nature of these accelerators, their lack of uniformity, and the fact that more will be added over time, invoking them from Wasm is problematic.
Extending the Veracruz system interface to incorporate accelerator-specific functionality would take us beyond Wasi, and require the use of support libraries for programming with Veracruz.
Instead, we opt for an interface built around \deffont{special files} in the Veracruz filesystem, with modules invoked by Wasm programs writing-to and reading-from these files, reusing existing programming idioms and filesystem support in Wasi.
\subsection{Threat model}
\label{subsect.veracruz.threat.model}

The Veracruz TCB includes the underlying isolate, the Veracruz runtime, and the implementation of the Veracruz proxy attestation service.
The host of the Veracruz attestation service must also be trusted by all parties, as must the native attestation services or keys in use.
The correctness of the various protocols in use---TLS, platform-specific native attestation, and PSA attestation---must also be trusted.

The Wasm execution engine must also be trusted to correctly execute a binary, so that a computation is faithfully executed according to the published bytecode semantics~\cite{DBLP:journals/cacm/RossbergTHSGWZB18, DBLP:conf/cpp/Watt18}, and that the program is unable to escape its sandbox, damage or spy on a delegate, or have any other side-effect than allowed by the Veracruz sandboxing model.
Recent techniques have been developed that use post-compilation verification to establish this trust~\cite{johnson2021}---we briefly discuss our ongoing experiments in this area in \S\ref{sect.closing.remarks}.
Compiler verification could be used to engender trust in the Wasm execution engine, though we are not aware of any verified, high-performance Wasm interpreters or JITs suitable for use with Veracruz at the time of writing (see~\cite{Watt2021Two} for progress toward this, however).
Memory issues have been implicated in attacks against isolates in the past~\cite{10.5555/3241189.3241231}---we write Veracruz in Rust in an attempt to avoid this, with the compiler therefore also trusted.

Veracruz does not defend against denial-of-service attacks: the delegate is in charge of scheduling execution, and liveness guarantees are therefore impossible to uphold.
A malicious principal can therefore deny others access to a computation's result, or refuse to provision a data input or program, thereby blocking the computation from even starting.

Different isolation technologies defend against different classes of attacker, and as Veracruz supports multiple technologies we must highlight these differences explicitly.

AWS Nitro Enclaves protect computations from the AWS customer running the EC2 instance associated with the isolate.
While AWS assures users that isolates are protected from employees and other insiders, these assurances are difficult to validate (and, as silicon manufacturer, AWS and its employees must always be trusted).
Our TCB therefore also contains the Nitro hardware, Linux host used inside the isolate, the attestation infrastructure for Nitro Enclaves, and any AWS insiders with access to that infrastructure.

For Arm CCA Realms only the \emph{Realm Management Monitor} (RMM, henceforth), a separation kernel isolating Realms from each other, has access to the memory of a Realm other than the software executing in the Realm itself.
Realms are protected from the non-secure hypervisor, and any other software running on the system other than the RMM, and will be protected against a class of physical attacks using memory encryption.
Our TCB therefore contains the RMM, the system hardware, Linux host inside the Realm, along with the attestation infrastructure for Arm CCA.

For IceCap our TCB includes the seL4 kernel which we rely on to securely isolate processes from one another, bolstered by a body of machine-checked proofs of the kernel's security and functional correctness (though at present these do not extend to the \texttt{EL2} configuration for AArch64).
For a typical hypervisor deployment of seL4, the SMMU is the only defence against physical attacks.

The TCB of Veracruz includes both local and remote stacks of hardware and software, while purely cryptographic techniques merely rely on a trustworthy implementation of a primitive and the correctness of the primitive itself.
As demonstrated in \S\ref{sect.evaluation}, Veracruz provides a degree of efficiency and practicality currently out of reach for purely cryptographic techniques, at the cost of this larger TCB.

Principals face a challenging class of threats stemming from collusion between the other principals, including the delegate.
Some algorithms may be particularly vulnerable to an unwanted declassification of secret inputs to any result receiver, and some attacks may be enhanced by collusion between principals---e.g., a side-channel inserted into the program for the benefit of the delegate.
As discussed in \S\ref{sect.hardware-backed.confidential.computing}, several powerful side-channel attacks have been demonstrated in the past against software executing within isolates, and other side-channels also exist including wall-clock execution time of the program, $\pi$, on the input data sets, and data sizes and arrival times leaked by TLS connections.
In cases where programs are secret, principals must trust the program provider not to collude with the result receiver, as a secret program could trivially intentionally leak data into the result or contain convert channels.
If the existence of this trust relationship is undesirable, then principals should insist on program declassification before enrolling in a computation.

\section{Evaluation}
\label{sect.evaluation}

This section uses the following test platforms: Intel Core i7-8700, \SI{16}{\gibi\byte} RAM, \SI{1}{\tera\byte} SSD (\coreiseven, henceforth); \texttt{c5.xlarge} AWS VM, \SI{8}{\gibi\byte} RAM, EBS (\ectwo, henceforth); Raspberry Pi 4, \SI{4}{\gibi\byte} RAM, \SI{32}{\giga\byte} $\mu$SD (\rpifour, henceforth).
We use GCC $9.30$ for x86-64, GCC $7.5.0$ for AArch64, and Wasi SDK-$14.0$ with LLVM $13.0$ for Wasm.

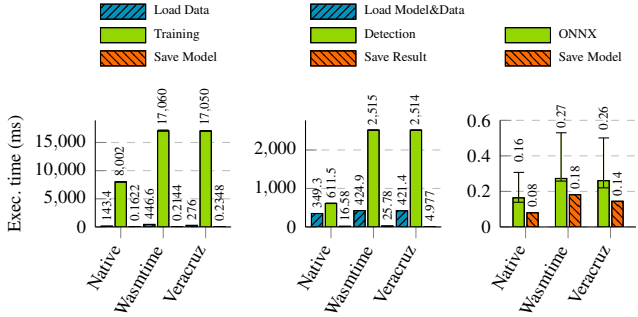
\begin{figure}[t]
    \vspace{-17pt}
    \begin{tikzpicture}
\begin{axis} [
    width=\columnwidth * 0.4, height=3cm, 
    ymajorgrids=true,
    grid style=dashed,
    align=center,
    ybar=1pt, 
    ymin=0.0,
    axis y line*=none, axis x line*=none,
    bar width = 0.15cm,
    xtick = data,
    symbolic x coords={Native, Wasmtime, Veracruz}, 
    x tick label style={rotate=60,anchor=east,font=\scriptsize},
    y tick label style={font=\scriptsize},
    nodes near coords,
    nodes near coords align={right},
    nodes near coords style={rotate=90, yshift=0.0cm, xshift=-0.05cm, font=\tiny, /pgf/number format/.cd,fixed relative,precision=4},
    point meta=rawy, 
    legend pos=north west,
    legend style={
        inner xsep=0pt,inner ysep=20pt,nodes={inner xsep=3pt},
        legend columns=1,
        anchor=south west,
        draw=none,
        fill=none,
        font=\tiny,
        },
    enlarge x limits={rel=0.3},
    legend cell align=left,
    xlabel style={yshift=0.0cm},
    ylabel style={yshift=-0.1cm, font=\scriptsize},
    scaled ticks=false,
    ylabel={Exec. time (ms)},
    area legend] 
]

\addplot [black,fill=arm-blue, postaction={pattern = north east lines}] table[x=platform, y expr=\thisrow{train-load}*1000] {evaluation/darknet.txt};
\addlegendentry[]{Load Data};
\addplot [black,fill=arm-green, postaction={pattern = none}, error bars/.cd, y dir=both, y explicit] table[x=platform, y expr=\thisrow{train-comp}*1000, y error minus expr=(\thisrow{train-comp}-\thisrow{train-l})*1000, y error plus expr=(\thisrow{train-u}-\thisrow{train-comp})*1000] {evaluation/darknet.txt};
\addlegendentry[]{Training};
\addplot [black,fill=arm-orange, postaction={pattern = north west lines}] table[x=platform, y expr=\thisrow{train-save}*1000] {evaluation/darknet.txt};
\addlegendentry[]{Save Model};
\end{axis}
\end{tikzpicture}
    \begin{tikzpicture}
\begin{axis} [
    width=\columnwidth * 0.4, height=3cm, 
    ymajorgrids=true,
    grid style=dashed,
    align=center,
    ybar=1pt, 
    ymin=0.0,
    axis y line*=none, axis x line*=none,
    bar width = 0.15cm,
    xtick = data,
    symbolic x coords={Native, Wasmtime, Veracruz}, 
    x tick label style={rotate=60,anchor=east,font=\scriptsize},
    y tick label style={font=\scriptsize},
    nodes near coords,
    nodes near coords align={right},
    nodes near coords style={rotate=90, yshift=0.0cm, xshift=-0.05cm, font=\tiny, /pgf/number format/.cd,fixed relative,precision=4},
    point meta=rawy, 
    legend pos=north west,
    legend style={
        inner xsep=0pt,inner ysep=20pt,nodes={inner xsep=3pt},
        legend columns=1,
        anchor=south west,
        draw=none,
        fill=none,
        font=\tiny,
        },
    enlarge x limits={rel=0.3},
    legend cell align=left,
    xlabel style={yshift=0.0cm},
    area legend] 
]

\addplot [black,fill=arm-blue, postaction={pattern = north east lines}] table[x=platform, y expr=(\thisrow{det-arg-load}+\thisrow{det-image-load})*1000] {evaluation/darknet.txt};
\addlegendentry[]{Load Model\&Data};
\addplot [black,fill=arm-green, postaction={pattern = none}, error bars/.cd, y dir=both, y explicit] table[x=platform, y expr=\thisrow{det-comp}*1000, y error minus expr=(\thisrow{det-comp}-\thisrow{det-l})*1000, y error plus expr=(\thisrow{det-u}-\thisrow{det-comp})*1000] {evaluation/darknet.txt};
\addlegendentry[]{Detection};
\addplot [black,fill=arm-orange, postaction={pattern = north west lines}] table[x=platform, y expr=\thisrow{box}*1000] {evaluation/darknet.txt};
\addlegendentry[]{Save Result};
\end{axis}
\end{tikzpicture}\hspace*{-10pt} 
    \begin{tikzpicture}
\begin{axis} [
    width=\columnwidth * 0.4, height=3cm, 
    ymajorgrids=true,
    grid style=dashed,
    align=center,
    ybar=1pt, 
    ymin=0.0,
    ymax=0.6,
    axis y line*=none, axis x line*=none,
    bar width = 0.15cm,
    xtick = data,
    symbolic x coords={Native, Wasmtime, Veracruz}, 
    x tick label style={rotate=60,anchor=east,font=\scriptsize},
    y tick label style={font=\scriptsize},
    nodes near coords,
    nodes near coords align={right},
    nodes near coords style={rotate=90, yshift=0.0cm, xshift=-0.05cm, font=\tiny, /pgf/number format/.cd,fixed relative,precision=2},
    point meta=rawy, 
    legend pos=north west,
    legend style={
        inner xsep=0pt,inner ysep=20pt,nodes={inner xsep=3pt},
        legend columns=1,
        anchor=south west,
        draw=none,
        fill=none,
        font=\tiny,
        },
    enlarge x limits={rel=0.3},
    legend cell align=left,
    xlabel style={yshift=0.0cm},
    area legend] 
]

\addplot [black,fill=arm-green, every node near coord/.append style={/tikz/xshift={ifthenelse(\coordindex<1,0.4cm,0.6cm)}}, postaction={pattern = none}, error bars/.cd, y dir=both, y explicit] table[x=platform, y expr=\thisrow{onnx-comp}*1000, y error minus expr=(\thisrow{onnx-comp}-\thisrow{onnx-l})*1000, y error plus expr=(\thisrow{onnx-u}-\thisrow{onnx-comp})*1000] {evaluation/darknet.txt} ;
\addlegendentry[]{ONNX};
\addplot [black,fill=arm-orange, postaction={pattern = north west lines}] table[x=platform, y expr=\thisrow{onnx-save}*1000] {evaluation/darknet.txt};
\addlegendentry[]{Save Model};
\end{axis}
\end{tikzpicture}
    \vspace{-10pt}
    \caption{Execution time of the DL examples, classifier training (L), inference (M), and \emph{ONNX} model aggregation (R)}
    \label{fig:veracruz.learning}
\end{figure}

\subsection{Case-study: deep learning}
\label{subsect.evaluation.case-study.deep.learning}

Training datasets, algorithms, and learnt models may be sensitive IP and the learning and inference processes are vulnerable to malicious changes in model parameters that can cause a negative influence on a model's behaviors that is hard to detect~\cite{bagdasaryan2020backdoor, mo2021ppfl}.
We present two Veracruz case-studies in protecting deep learning (DL henceforth) applications: privacy-preserving training and inference, and privacy-preserving model aggregation service, a step toward \emph{federated} DL.
We use \emph{Darknet}~\cite{darknet13,mo2020darknetz} in both cases, and the \emph{Open Neural Network eXchange}~\cite{onnx2019, connxr2021} (\emph{ONNX}, henceforth) as the aggregation format.
We focus on the \emph{execution time} of training, inference, and model aggregation on the \coreiseven~test platform.

In the training and inference case-study, the program receives input datasets from the respective data providers and a pre-learnt model from a model provider.
Thereafter, the provisioned program starts training or inference, protected inside Veracruz.
The results---that is, the trained model or prediction---are made available to a result receiver.
In the model aggregation case-study, clients conduct \emph{local training} with their favorite DL frameworks, convert the models to \emph{ONNX} format, and provision these derived models into Veracruz.
The program then aggregates the models, making the result available to all clients.
By converting to \emph{ONNX} locally, we support a broad range of local training frameworks---i.e., \emph{PyTorch}~\cite{pytorch2019}, \emph{Tensorflow}~\cite{tensorflow2015}, \emph{Darknet}, or similar.


We trained a LeNet\cite{lecun1998gradient} on \emph{MNIST}\cite{lecun1998gradient}, a dataset of handwritten digits consisting of $60,000$ training and $10,000$ validation images.
Each image is $28{\times}28$ pixels and less than \SI{1}{\kibi\byte}; we used a batch size of $100$ in training, obtaining a trained model of \SI{186}{\kibi\byte}.
We take the average of $20$ trials for training on 100 $batches$ (hence, 10,000 images) and then ran inference on one image.
For aggregation, we use three copies of this \emph{Darknet} model (\SI{186}{\kibi\byte}), obtaining three \emph{ONNX} models (\SI{26}{\kibi\byte}), performing $200$ trials for aggregation, as aggregation time is significantly less.
Results are presented in Fig.~\ref{fig:veracruz.learning}.

For all DL tasks we observe the same execution time between Wasmtime and Veracruz, as expected, with both around $2.1$--$4.1\times$ slower than native CPU-only execution, likely due to more aggressive code optimization available in native compilers. 
However, the similarity between Wasmtime and Veracruz diverges for file operations such as loading and saving of model data.
Loading data from disk is $1.2$--$3.1\times$ slower when using Wasmtime compared to executing natively.
However, I/O in Veracruz is usually \emph{faster} than Wasmtime, and sometimes faster than native execution, e.g., when saving images in inference.
This is likely due to Veracruz's in-memory filesystem exhibiting a faster read and write speed transferring data, compared to the SSD of the test machine.



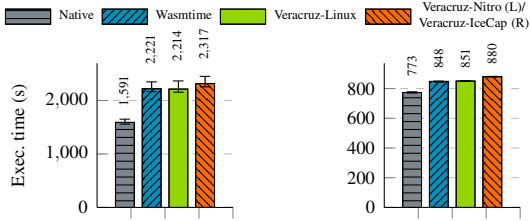
\begin{figure}[t]
\centering
\begin{tikzpicture}

\begin{axis} [
    width=\columnwidth*0.4, height=3.5cm, 
    ymajorgrids=true,
    grid style=dashed,
    align=center,
    ybar=3pt, 
    axis y line*=none, axis x line*=none,
    bar width = 0.25cm,
    ymin=0,
    xtick = {}, 
    symbolic x coords={0}, 
    xticklabels = {}, 
    x tick label style={rotate=60,anchor=east,font=\scriptsize},
    y tick label style={font=\scriptsize},
    nodes near coords,
    nodes near coords align={right},
    nodes near coords style={rotate=90, yshift=0.0cm, xshift=0cm, font=\tiny, /pgf/number format/.cd,fixed zerofill,precision=0},
    point meta=rawy, 
    legend style={
        inner xsep=0pt,inner ysep=0pt,nodes={inner xsep=3pt},
        yshift=10pt,
        xshift=40pt,
        legend columns=4,
        anchor=south,
        draw=none,
        font=\tiny,
        fill=none,
        at= {(current axis.north)},
        },
    legend cell align=left,
    xlabel style={yshift=0.0cm},
    ylabel style = {font=\scriptsize},
    ylabel={Exec.~time (s)},
    area legend] 
]

\addplot [black,fill=arm-grey, every node near coord/.append style={/tikz/xshift=0.1cm}, postaction={pattern = horizontal lines}, error bars/.cd, y dir=both, y explicit] table[x=totalTime,y expr=\thisrow{native}, y error minus=native-min, y error plus=native-max] {evaluation/ec2-vod_total_time.txt};
\addlegendentry[]{Native};
\addplot [black,fill=arm-blue, every node near coord/.append style={/tikz/xshift=0.2cm}, postaction={pattern = north east lines}, error bars/.cd, y dir=both, y explicit] table[x=totalTime,y expr=\thisrow{wasmtime}, y error minus=wasmtime-min, y error plus=wasmtime-max] {evaluation/ec2-vod_total_time.txt};
\addlegendentry[]{Wasmtime};
\addplot [black,fill=arm-green, every node near coord/.append style={/tikz/xshift=0.25cm}, postaction={pattern = none}, error bars/.cd, y dir=both, y explicit] table[x=totalTime,y expr=\thisrow{vc-linux}, y error minus=vc-linux-min, y error plus=vc-linux-max] {evaluation/ec2-vod_total_time.txt};
\addlegendentry[]{Veracruz-Linux};
\addplot [black,fill=arm-orange, every node near coord/.append style={/tikz/xshift=0.2cm}, postaction={pattern = north west lines}, error bars/.cd, y dir=both, y explicit] table[x=totalTime,y expr=\thisrow{vc-nitro}, y error minus=vc-nitro-min, y error plus=vc-nitro-max] {evaluation/ec2-vod_total_time.txt};
\addlegendentry[]{Veracruz-Nitro (L)/\\Veracruz-IceCap (R)};

\end{axis}
\end{tikzpicture}\hspace*{-80pt}
\begin{tikzpicture}

\begin{axis} [
    width=\columnwidth*0.4, height=3.5cm, 
    ymajorgrids=true,
    grid style=dashed,
    align=center,
    ybar=3pt, 
    axis y line*=none, axis x line*=none,
    bar width = 0.25cm,
    ymin=0,
    xtick = {}, 
    symbolic x coords={0}, 
    xticklabels = {}, 
    x tick label style={rotate=60,anchor=east,font=\scriptsize},
    y tick label style={font=\scriptsize},
    legend style={
        inner xsep=0pt,inner ysep=10pt,nodes={inner xsep=3pt},
        yshift=10pt,
        legend columns=4,
        anchor=south,
        draw=none,
        font=\tiny,
        fill=none,
        at= {(current axis.north)},
        },
    legend cell align=left,
    nodes near coords,
    nodes near coords align={right},
    nodes near coords style={rotate=90, yshift=0.0cm, xshift=0.0cm, font=\tiny, /pgf/number format/.cd,fixed zerofill,precision=0},
    point meta=rawy, 
    xlabel style={yshift=0.0cm},
    ylabel style = {font=\scriptsize},
    ylabel={},
    ] 
]

\addplot [black,fill=arm-grey, every node near coord/.append style={/tikz/xshift=0.1cm}, postaction={pattern = horizontal lines}, error bars/.cd, y dir=both, y explicit] table[x=totalTime,y expr=\thisrow{native}, y error minus=native-min, y error plus=native-max] {evaluation/rpi4-vod.txt};
\addplot [black,fill=arm-blue, every node near coord/.append style={/tikz/xshift=0.05cm}, postaction={pattern = north east lines}, error bars/.cd, y dir=both, y explicit] table[x=totalTime,y expr=\thisrow{wasmtime}, y error minus=wasmtime-min, y error plus=wasmtime-max] {evaluation/rpi4-vod.txt};
\addplot [black,fill=arm-green, every node near coord/.append style={/tikz/xshift=0.05cm}, postaction={pattern = none}, error bars/.cd, y dir=both, y explicit] table[x=totalTime,y expr=\thisrow{vc-linux}, y error minus=vc-linux-min, y error plus=vc-linux-max] {evaluation/rpi4-vod.txt};
\addplot [black,fill=arm-orange, every node near coord/.append style={/tikz/xshift=0.05cm}, postaction={pattern = north west lines}, error bars/.cd, y dir=both, y explicit] table[x=totalTime,y expr=\thisrow{vc-icecap}, y error minus=vc-icecap-min, y error plus=vc-icecap-max] {evaluation/rpi4-vod.txt};
\end{axis}

\end{tikzpicture}
\spacehack{10pt}
\caption{Video object detection execution time on \ectwo~(L) and \rpifour~(R)}
\label{fig:veracruz.vod}
\end{figure}

\subsection{Case-study: video object detection}
\label{subsect.evaluation.case-study.video.object.detection}

We have used Veracruz to prototype a \emph{Confidential FaaS}, running on AWS Nitro Enclaves and using \emph{Kubernetes}~\cite{kubernetes}.
In this model, a cloud infrastructure or other delegate initializes an isolate containing only the Veracruz runtime and provides an appropriate global policy file.
Confidential functions are registered in a \emph{Confidential Computing as a Service} (CCFaaS, henceforth) component, which acts as a registry for clients wishing to use the service and which collaborates, on behalf of clients, with a \emph{Veracruz as a Service} (VaaS, henceforth) component which manages the lifetime of any spawned Veracruz instances.
Together, the CCFaaS and VaaS components draft policies and initialize Veracruz instances, while attestation is handled by clients, using the proxy attestation service.

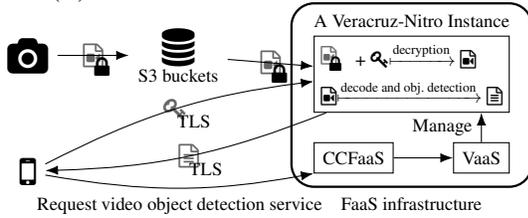
\begin{figure}[t!]
    \centering
    \vspace*{-10pt}
    \begin{tikzpicture}[
    singlearrow/.style={-Latex},
    doublearrow/.style={Latex-Latex},
    lock/.style={xshift=-4, yshift=4},
    tls/.style={xshift=-1,yshift=3},
    font=\scriptsize
]
\node (camera) {\Large\faCamera};
\node[align=center] (s3) at ($(camera) + (2,0)$) {\Large\faDatabase\\S3 buckets};
\node[opacity=0.6] (video) at ($(s3) + (2,0)$) {\faFileVideoO};
\node[anchor=west] (key) at (video.east) {+ \faKey \(\xmapsto{\text{decryption}}\) \faFileVideoO};
\node[anchor=west] (compute) at ($(video.west) + (0,-0.5)$) {\faFileVideoO \( \xmapsto{\text{decode and obj. detection}}\) \faFileTextO};
\node[draw, fit=(video)(key)(compute), inner sep=-1pt] (veracruz) {};
\node[anchor=south, inner ysep=-1pt, yshift=4pt] (veracruz-label) at (veracruz.north) {A Veracruz-Nitro Instance};
\node[anchor=west, align=center, draw] (ccfaas) at ($(veracruz.south west) + (0,-0.6)$) {CCFaaS};
\node[anchor=east, align=center, draw] (vaas) at ($(veracruz.south east) + (0,-0.6)$) {VaaS};
\node (phone) at ($(camera) + (0,-1.55)$) {\Large\faMobilePhone};

\draw[singlearrow] (camera) to node[pos=0.5, opacity=0.6] (camera-video) {\small\faFileVideoO} (s3);
\draw[singlearrow] (s3) to node[pos=0.5, opacity=0.6] (iotex-video) {\small\faFileVideoO} (veracruz);
\draw[singlearrow] (phone) to[bend left=10] node[pos=0.5, opacity=0.6] (phone-key) {\small\faKey} (veracruz);
\draw[singlearrow] (veracruz) to[bend left=10] node[pos=0.5, opacity=0.6] (veracruz-document) {\small\faFileTextO} (phone);
\draw[singlearrow] (phone) to[bend right=10] node[pos=0.5,below] {Request video object detection service} (ccfaas.south west);
\draw[singlearrow] (ccfaas) to[bend right=0] node[pos=0.5,below] {} (vaas);
\draw[singlearrow] (vaas) to[bend right=0] node[pos=0.5,left] (vaas-manage) {Manage} (vaas |- veracruz.south);

\node[lock] at (camera-video.south east) {\small\faLock};
\node[lock] at (iotex-video.south east) {\small\faLock};
\node[lock] at (video.south east) {\scriptsize\faLock};
\node[tls] at (phone-key.south east) {TLS};
\node[tls] at (veracruz-document.south east) {TLS};

\node[fit=(veracruz)(ccfaas)(vaas)(vaas-manage)(veracruz-label), draw, label={below:FaaS infrastructure}, line width=0.30mm, inner sep=5pt, rounded corners=3mm] (faas) {};
\end{tikzpicture}%
    \caption{Video object detection case-study}
    \label{fig:APP-model-2}
\end{figure}

Building atop this confidential FaaS infrastructure, we applied Veracruz in a full end-to-end encrypted video object detection flow (see Fig.~\ref{fig:APP-model-2}).
Our intent is to demonstrate that Veracruz can be applied to industrially-relevant use-cases: here, a video camera manufacturer wishes to offer an object detection service to their customers while providing believable guarantees that they cannot access customer video.

The encrypted video clips originating from an IoTeX Ucam video camera~\cite{iotexucam} are stored in an \emph{AWS S3} bucket.
The encryption key is owned by the camera operator and perhaps generated by client software on their mobile phone or tablet.
Independently, a video processing and object detection function, compiled to Wasm, is registered with the CCFaaS component which takes on the role of program provider in the Veracruz computation.
This function makes use of the Cisco \texttt{openh264} library as well as the Darknet neural network framework and a prebuilt YOLOv3 model, as previously discussed in \S\ref{subsect.evaluation.case-study.deep.learning}, for object detection (our support for Wasi eased this porting).

Upon the request of the camera owner, the CCFaaS and VaaS infrastructure spawn a new AWS Nitro Enclave loaded with the Veracruz runtime, and configured using an appropriate global policy that lists the camera owner as having the role of data provider and result receiver.
The confidential FaaS infrastructure forwards the global policy to the camera owner, where it is automatically analyzed by their client software, with the camera owner thereafter attesting the AWS Nitro Enclave instance.
If the global policy is acceptable, and attestation succeeds, the camera owner securely connects to the spawned isolate, containing the Veracruz runtime, and securely provisions their decryption key using TLS in their role as data provider.
The encrypted video clip is also then provisioned into the isolate, by a dedicated AWS S3 application, which is also listed in the global policy as a data provider, and the computation can then go ahead.
Once complete, meta-data containing the bounding boxes of any object detected in the frames of the video clips can be securely retrieved by the camera owner via TLS, in their result receiver role, for interpretation by their client software.

Note that in this FaaS infrastructure desirable cloud application characteristics are preserved: the computation is on-demand and scaleable, and our infrastructure allows multiple instances of Veracruz, running different functions, to be executed concurrently.
Only the AWS S3 application, the camera owner's client application and the video decoding and object detection function are specific to this use-case.
All other modules are generic, allowing other applications to be implemented.
Moreover, note that no user credentials or passwords are shared directly with the FaaS infrastructure in realizing this flow, beyond the name of the video clip to retrieve from the AWS S3 bucket and a one-time access credential for the AWS S3 application.
Decryption keys are only shared with the Veracruz runtime inside an attested isolate.

We benchmark by passing a $1920{\times}1080$ video to the object detection program, which decodes frame by frame, converts, downscales, and passes frames to the ML model.
We compare four configurations on two different platforms:
\begin{itemize*}
\item
On \ectwo, a native \texttt{x86-64} binary on Amazon Linux; a Wasm binary under Wasmtime-0.27; a Wasm binary inside Veracruz as a Linux process; a Wasm binary inside Veracruz on AWS Nitro Enclaves.
The video is $240$ frames long and fed to the YOLOv3-608 model~\cite{yolov3}. 
\item
On \rpifour: a native AArch64 binary on Ubuntu 18.04 Linux; a Wasm binary under Wasmtime-0.27; a Wasm binary inside Veracruz as a Linux process; a Wasm binary inside Veracruz on IceCap.
Due to memory limits the video is $240$ frames long and fed to the YOLOv3-tiny model~\cite{yolov3}.
\end{itemize*}
We take the native \texttt{x86-64} configuration as our baseline, and present average runtimes for each configuration, along with observed extremes, in Fig.~\ref{fig:veracruz.vod}.

\begin{table}[t]
    \centering
    \begin{tabular}{ll}
    \hline
    Description & Time (ms) \\
      \hline
      \emph{Proxy Attestation Service start} & 7 \\
      \emph{Onboard new Veracruz isolate} & 3122 \\
      \emph{Request attestation message} & 54 \\
      \emph{Initialization of Veracruz isolate} & 1 \\
      \emph{Check hashes (including TLS handshake)} & 184 \\
      \emph{Provision object detection program} & 798 \\
      \emph{Provision data (model, video)} & 282323 \\
      \hline
    \end{tabular}
    \caption{Breakdown of Veracruz deployment overheads for the video object detection use-case on AWS Nitro Enclaves}
    \label{fig.attestation.overheads}
\end{table}

\paragraph{\ectwo~results}
Wasm (with experimental SIMD support in Wasmtime) has an overhead of ${\sim}39$\% over native code; most CPU cycles are spent in matrix multiplication, which the native compiler can better autovectorize than the Wasm compiler.
The vast majority of execution time is spent in neural network inference, rather than video decode or image downscaling.
Since execution time is dominated by the Wasm execution, Veracruz overhead is negligible.
A ${\sim}5$\% performance discrepancy exists between Nitro and Wasmtime, which could originate from our observation that Nitro is slower at loading data into an enclave, but faster at writing, though Nitro runs a different kernel with a different configuration, on a separate CPU, making this hard to pinpoint.
Deployment overheads for Nitro are presented in Table~\ref{fig.attestation.overheads}, showing a breakdown of overheads for provisioning a new Veracruz instance.

\paragraph{\rpifour~results}
The smaller ML model significantly improves inference performance at the expense of accuracy.
Wasm has an overhead of ${\sim}10$\% over native code, smaller than the gap on \ectwo, and could be due to reduced vectorization support in GCC's AArch64 backend.
Veracruz overhead is again negligible, though IceCap induces an overhead of ${\sim}3$\% over Veracruz-Linux.
This observation approximately matches the overhead of ${\sim}2$\% for CPU-bound workloads measured in Fig.~\ref{fig.icecap.overheads}, explained by extra context switching through trusted resource management services during scheduling operations.

Using ``native modules'', introduced in \S\ref{subsect.veracruz.ad.hoc.acceleration}, explicit support for neural network inference could be added to the Veracruz runtime, though our results above suggest a max ${\sim}38$\% performance boost by pursuing this, likely less due to the costs of marshalling data between the native module and Veracruz file system.
For larger performance boosts, dedicated ML acceleration could be used, requiring support from the Veracruz runtime, though establishing trust in accelerators outside the isolate is hard, with PCIe attestation still a work-in-progress.


\subsection{Further comparisons}

\begin{figure}[t!]
\centering
\vspace*{-4pt}
\hspace*{-10pt}
\begin{tikzpicture} [baseline]
\begin{axis} [
    width=\columnwidth * 1.08, height=3cm, 
    ymajorgrids=true,
    grid style=dashed,
    align=center,
    ybar=0.5pt, 
    xmin=2mm,
    xmax=trmm,
    ymin=0.75,
    ymax=2.375,
    axis y line*=none, axis x line*=none,
    bar width = 0.06cm,
    xtick = data,
    symbolic x coords={2mm, 3mm, adi, atax, bicg, cholesky, correlation, covariance, deriche, doitgen, durbin, fdtd-2d, floyd-warshall, gemm, gemver, gesummv, gramschmidt, heat-3d, jacobi-1d, jacobi-2d, lu, ludcmp, mvt, nussinov, seidel-2d, symm, syr2k, syrk, trisolv, trmm}, 
    x tick label style={rotate=60,anchor=east,font=\scriptsize},
    y tick label style={font=\scriptsize},
    yticklabel={$\pgfmathprintnumber{\tick}\times$},
    xtick distance=-5pt,
    point meta=rawy, 
    legend style={
        inner xsep=0pt,inner ysep=10pt,nodes={inner xsep=3pt},
        legend columns=3,
        anchor=south east,
        draw=none,
        font=\tiny,
        },
    legend cell align=left,
    enlarge x limits=0.02,
    xlabel style={yshift=0.0cm},
    ylabel style={yshift=-0.1cm, font=\scriptsize}, 
    area legend]
]

\addplot [black,fill=arm-blue, postaction={pattern = north east lines}] table[x=prog,y expr=\thisrow{wasmtime}/\thisrow{native}] {evaluation/ec2-polybench.txt};
\addlegendentry[]{Wasmtime};
\addplot [black,fill=arm-green, postaction={pattern = none}] table[x=prog,y expr=\thisrow{linux}/\thisrow{native}] {evaluation/ec2-polybench.txt};
\addlegendentry[]{Veracruz-Linux};
\addplot [black,fill=arm-orange, postaction={pattern = north west lines}] table[x=prog,y expr=\thisrow{nitro}/\thisrow{native}] {evaluation/ec2-polybench.txt};
\addlegendentry[]{Veracruz-Nitro};
\end{axis}
\end{tikzpicture}\hspace*{-14pt}
\begin{tikzpicture} [baseline]
\begin{axis} [
    width=\columnwidth * 0.225, height=3cm, 
    ymajorgrids=true,
    grid style=dashed,
    align=center,
    ybar=0.5pt, 
    xmin=gmean,
    xmax=gmean,
    ymin=0.75,
    ymax=2.375,
    yticklabels=false,
    axis y line*=none, axis x line*=none,
    bar width = 0.06cm,
    xtick = data,
    symbolic x coords={gmean}, 
    x tick label
    style={rotate=60,anchor=east,font=\scriptsize},
    y tick label style={font=\scriptsize},
    xtick distance=-5pt,
    point meta=rawy, 
    legend style={
        inner xsep=0pt,inner ysep=10pt,nodes={inner xsep=3pt},
        legend columns=3,
        anchor=south east,
        draw=none,
        font=\tiny,
        },
    legend cell align=left,
    enlarge x limits=0.02,
    xlabel style={yshift=0.0cm},
    ylabel style={yshift=-0.1cm, font=\scriptsize}, 
    area legend] 
]

\addplot [black,fill=arm-blue, postaction={pattern = north east lines}] table[x=prog,y expr=\thisrow{wasmtime}/\thisrow{native}] {evaluation/ec2-polybench-gmean.txt};
\addplot [black,fill=arm-green, postaction={pattern = none}] table[x=prog,y expr=\thisrow{linux}/\thisrow{native}] {evaluation/ec2-polybench-gmean.txt};
\addplot [black,fill=arm-orange, postaction={pattern = north west lines}] table[x=prog,y expr=\thisrow{nitro}/\thisrow{native}] {evaluation/ec2-polybench-gmean.txt};
\end{axis}
\end{tikzpicture}
\spacehack{10pt}
\caption{Relative execution time (vs.~native) of PolyBench/C (large dataset) on \ectwo.
\texttt{gmean} shows the geometric mean of all results}
\label{fig:veracruz-polybench}
\end{figure}
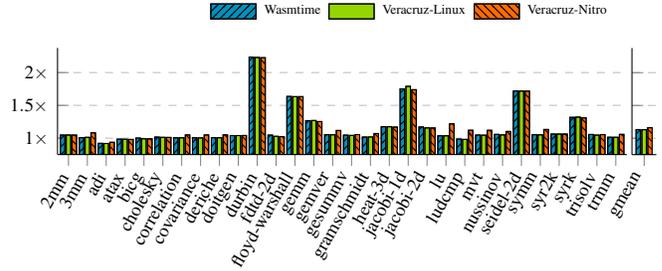

\paragraph{PolyBench/C microbenchmarks}
We further evaluate the performance of Veracruz on compute-bound programs using the PolyBench/C suite (version 4.2.1-beta)~\cite{polybench}, a suite of small, simple computationally-intensive kernels.
We compare execution time of four different configurations on the \ectwo~ instance running \emph{Amazon Linux 2}: a native \texttt{x86-64} binary; a Wasm binary under Wasmtime-0.27; a Wasm binary under Veracruz as a Linux process; and a Wasm binary executing under Veracruz in an AWS Nitro Enclave.
We take \texttt{x86-64} as our baseline, and present results in Fig.~\ref{fig:veracruz-polybench}.
Wasmtime's overhead against native CPU execution is relatively small with a geometric mean of ${\sim}13$\%, though we observe that some test programs execute even faster under Wasmtime than when natively compiled.
Again, we compile our test programs with Wasmtime's experimental support for SIMD proposal, though this boosts performance for only a few programs.
Veracruz-Linux doesn't exhibit a visible overhead compared to Wasmtime, which is expected as most execution time is spent in Wasmtime, and the presence of the Veracruz VFS is largely irrelevant for CPU-bound programs.
Veracruz-Nitro exhibits a small but noticeable overhead (${\sim}3$\%) compared to Veracruz-Linux, likely due to the reasons mentioned in \S\ref{subsect.evaluation.case-study.video.object.detection}.

\paragraph{VFS performance}

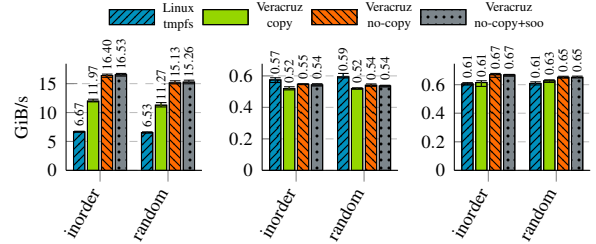
\begin{figure}[t!]
\centering
\newcommand{\gibs}[1]{((67108864/\thisrow{#1})/1024/1024/1024)}
\vspace*{-10pt}
\begin{tikzpicture}
\begin{axis} [
    width=\columnwidth * 0.4, height=3cm, 
    ymajorgrids=true,
    grid style=dashed,
    align=center,
    ybar=1pt, 
    ymin=0,
    axis y line*=none, axis x line*=none,
    bar width = 0.15cm,
    xtick = data,
    symbolic x coords={
        inorder,
        random,
    }, 
    x tick label style={rotate=60,anchor=east,font=\scriptsize},
    y tick label style={font=\scriptsize},
    nodes near coords,
    nodes near coords align={right},
    nodes near coords style={rotate=90, yshift=0.0cm, xshift=-0.05cm, font=\tiny, /pgf/number format/.cd,fixed zerofill,precision=2},
    point meta=rawy, 
    legend pos=north west,
    legend style={
        inner xsep=0pt,inner ysep=10pt,nodes={inner xsep=3pt},
        legend columns=1,
        anchor=south west,
        draw=none,
        fill=none,
        font=\tiny,
        },
    enlarge x limits={rel=0.5},
    legend cell align=left,
    xlabel style={yshift=0.0cm},
    ylabel style={yshift=-0.1cm, font=\scriptsize},
    ylabel={GiB/s},
    area legend] 
]

\addplot [black,fill=arm-blue, postaction={pattern = north east lines}, error bars/.cd, y dir=both, y explicit] table[skip first n=1, x=name, y expr=\gibs{r-wt}, y error minus expr=\gibs{r-wt}-\gibs{r-wt-l}, y error plus expr=\gibs{r-wt-u}-\gibs{r-wt}] {evaluation/fs.txt}; 
\addplot [black,fill=arm-green, postaction={pattern = none}, error bars/.cd, y dir=both, y explicit] table[skip first n=1, x=name, y expr=\gibs{r-v-copy}, y error minus expr=\gibs{r-v-copy}-\gibs{r-v-copy-l}, y error plus expr=\gibs{r-v-copy-u}-\gibs{r-v-copy}] {evaluation/fs.txt};
\addplot [black,fill=arm-orange, postaction={pattern = north west lines}, error bars/.cd, y dir=both, y explicit] table[skip first n=1, x=name, y expr=\gibs{r-v-no}, y error minus expr=\gibs{r-v-no}-\gibs{r-v-no-l}, y error plus expr=\gibs{r-v-no-u}-\gibs{r-v-no}] {evaluation/fs.txt};
\addplot [black,fill=arm-grey, postaction={pattern = dots}, error bars/.cd, y dir=both, y explicit] table[skip first n=1, x=name, y expr=\gibs{r-soo}, y error minus expr=\gibs{r-soo}-\gibs{r-soo-l}, y error plus expr=\gibs{r-soo-u}-\gibs{r-soo}] {evaluation/fs.txt}; 
\end{axis}

\end{tikzpicture}\hspace*{-38pt}
\begin{tikzpicture}
\begin{axis} [
    width=\columnwidth * 0.4, height=3cm, 
    ymajorgrids=true,
    grid style=dashed,
    align=center,
    ybar=1pt, 
    ymin=0,
    axis y line*=none, axis x line*=none,
    bar width = 0.15cm,
    xtick = data,
    symbolic x coords={
        inorder,
        random,
    }, 
    x tick label style={rotate=60,anchor=east,font=\scriptsize},
    y tick label style={font=\scriptsize},
    nodes near coords,
    nodes near coords align={right},
    nodes near coords style={rotate=90, yshift=0.0cm, xshift=-0.05cm, font=\tiny, /pgf/number format/.cd,fixed zerofill,precision=2},
    point meta=rawy, 
    legend pos=north west,
    legend style={
        inner xsep=0pt,inner ysep=10pt,nodes={inner xsep=3pt},
        legend columns=4,
        anchor=south,
        draw=none,
        fill=none,
        font=\tiny,
        at= {(current axis.north)}
        },
    enlarge x limits={rel=0.5},
    legend cell align=left,
    xlabel style={yshift=0.0cm},
    ylabel style={yshift=-0.1cm, font=\scriptsize},
    area legend] 
]

\addplot [black,fill=arm-blue, postaction={pattern = north east lines}, error bars/.cd, y dir=both, y explicit] table[skip first n=1, x=name, y expr=\gibs{w-wt}, y error minus expr=\gibs{w-wt}-\gibs{w-wt-l}, y error plus expr=\gibs{w-wt-u}-\gibs{w-wt}] {evaluation/fs.txt}; 
\addlegendentry[]{Linux\\tmpfs};
\addplot [black,fill=arm-green, postaction={pattern = none}, error bars/.cd, y dir=both, y explicit] table[skip first n=1, x=name, y expr=\gibs{w-v-copy}, y error minus expr=\gibs{w-v-copy}-\gibs{w-v-copy-l}, y error plus expr=\gibs{w-v-copy-u}-\gibs{w-v-copy}] {evaluation/fs.txt};
\addlegendentry[]{Veracruz\\copy};
\addplot [black,fill=arm-orange, postaction={pattern = north west lines}, error bars/.cd, y dir=both, y explicit] table[skip first n=1, x=name, y expr=\gibs{w-v-no}, y error minus expr=\gibs{w-v-no}-\gibs{w-v-no-l}, y error plus expr=\gibs{w-v-no-u}-\gibs{w-v-no}] {evaluation/fs.txt};
\addlegendentry[]{Veracruz\\no-copy};
\addplot [black,fill=arm-grey, postaction={pattern = dots}, error bars/.cd, y dir=both, y explicit] table[skip first n=1, x=name, y expr=\gibs{w-soo}, y error minus expr=\gibs{w-soo}-\gibs{w-soo-l}, y error plus expr=\gibs{w-soo-u}-\gibs{w-soo}] {evaluation/fs.txt}; 
\addlegendentry[]{Veracruz\\no-copy+soo};
\end{axis}
\end{tikzpicture}\hspace*{-55pt}
\begin{tikzpicture}
\begin{axis} [
    width=\columnwidth * 0.4, height=3cm, 
    ymajorgrids=true,
    grid style=dashed,
    align=center,
    ybar=1pt, 
    ymin=0,
    axis y line*=none, axis x line*=none,
    bar width = 0.15cm,
    xtick = data,
    symbolic x coords={
        inorder,
        random,
    }, 
    x tick label style={rotate=60,anchor=east,font=\scriptsize},
    y tick label style={font=\scriptsize},
    nodes near coords,
    nodes near coords align={right},
    nodes near coords style={rotate=90, yshift=0.0cm, xshift=-0.05cm, font=\tiny, /pgf/number format/.cd,fixed zerofill,precision=2},
    point meta=rawy, 
    legend pos=north west,
    legend style={
        inner xsep=0pt,inner ysep=10pt,nodes={inner xsep=3pt},
        legend columns=1,
        anchor=south west,
        draw=none,
        fill=none,
        font=\tiny,
        },
    enlarge x limits={rel=0.5},
    legend cell align=left,
    xlabel style={yshift=0.0cm},
    ylabel style={yshift=-0.1cm, font=\scriptsize},
    area legend] 
]

\addplot [black,fill=arm-blue, postaction={pattern = north east lines}, error bars/.cd, y dir=both, y explicit] table[skip first n=1, x=name, y expr=\gibs{u-wt}, y error minus expr=\gibs{u-wt}-\gibs{u-wt-l}, y error plus expr=\gibs{u-wt-u}-\gibs{u-wt}] {evaluation/fs.txt}; 
\addplot [black,fill=arm-green, postaction={pattern = none}, error bars/.cd, y dir=both, y explicit] table[skip first n=1, x=name, y expr=\gibs{u-v-copy}, y error minus expr=\gibs{u-v-copy}-\gibs{u-v-copy-l}, y error plus expr=\gibs{u-v-copy-u}-\gibs{u-v-copy}] {evaluation/fs.txt};
\addplot [black,fill=arm-orange, postaction={pattern = north west lines}, error bars/.cd, y dir=both, y explicit] table[skip first n=1, x=name, y expr=\gibs{u-v-no}, y error minus expr=\gibs{u-v-no}-\gibs{u-v-no-l}, y error plus expr=\gibs{u-v-no-u}-\gibs{u-v-no}] {evaluation/fs.txt};
\addplot [black,fill=arm-grey, postaction={pattern = dots}, error bars/.cd, y dir=both, y explicit] table[skip first n=1, x=name, y expr=\gibs{u-soo}, y error minus expr=\gibs{u-soo}-\gibs{u-soo-l}, y error plus expr=\gibs{u-soo-u}-\gibs{u-soo}] {evaluation/fs.txt}; 
\end{axis}

\end{tikzpicture}
\spacehack{10pt}
\caption{VFS bandwidth: read (L), write (M) and update (R)}
\label{fig:veracruz-vfs}
\end{figure}

We evaluate Veracruz VFS I/O performance, previously discussed in \S\ref{subsect.veracruz.programming.model}.
Performance is measured by timing common granular file-system operations and dividing by input size, to find the expected bandwidth.

Results gathered on \coreiseven~test platform with a swap size of zero so that measurements would not be invalidated by physical disk access, are presented in Fig.~\ref{fig:veracruz-vfs}.
Here, \deffont{read} denotes bandwidth of file read operations, \deffont{write} denotes bandwidth of file write operations with no initial file, and \deffont{update} denotes bandwidth of file write operations with an existing file. 
We use two access patterns, in-order and random, to avoid measuring only file-system-friendly access patterns.
All random inputs, for both data and access patterns, used reproducible, pseudorandom data generated by \texttt{xorshift64} to ensure consistency between runs.
All operations manipulate a \SI{64}{\mebi\byte} file with \SI{16}{\kibi\byte} buffer size---in practice, we expect most files will be within an order of magnitude of this size.

We compare variations of our VFS against Linux's \texttt{tmpfs}, the standard in-memory filesystem for Linux.
\deffont{Veracruz copy} moves data between the Wasm's sandboxed memory and the VFS through two copies, one at the Wasi API layer, and one at the internal VFS API layer.
\deffont{Veracruz no-copy} improved on this by performing a single copy directly from the Wasm's sandboxed memory into the destination in the VFS.
This was made possible thanks to Rust's borrow checker, which is able to express the temporarily shared ownership of the Wasm's sandboxed memory without sacrificing memory or lifetime safety.
In theory this overhead can be reduced to zero copies through \texttt{memmap}, however this API is not available in standard Wasi.
\deffont{Veracruz no-copy+soo} is our latest design, extending the no-copy implementation with a small-object optimization (SOO) \texttt{iovec} implementation---a Wasi structure describing a set of buffers containing data to be operated on, which for the majority of operations contain a reference to a single buffer.
Through this, we inline two or fewer buffers into the \texttt{iovec} structure itself, completely removing memory allocations from the read and write path for all programs we tested with.
Performance impact is negligible, however.

Being in an-memory filesystem, the internal representation is relatively simple: directories and a global \texttt{inode} table are implemented using hash tables, with each file represented as a vector of bytes.
While apparently na\"{i}ve, these data-structures have seen decades of optimization for in-memory performance, and even sparse files perform efficiently due to RAM over-commitment by the runtimes.
However, we were still surprised to see very close performance between Veracruz and \texttt{tmpfs}, with Veracruz nearly doubling the \texttt{tmpfs} performance for reads, likely due to the overhead of kernel syscalls necessary to communicate with \texttt{tmpfs} in Linux.
(Unfortunately \texttt{tmpfs} is deeply integrated into the Linux VFS layer, so it is not possible to compare with \texttt{tmpfs} in isolation.)

Both Veracruz and \texttt{tmpfs} use hash tables to store directory information, with the file data-structure and memory allocator representing significant differences.
In Veracruz we use byte vectors backed by the runtime's general purpose allocator, whereas \texttt{tmpfs} uses a tree of pages backed by the Linux VFS's page cache, which acts as a cache-aware fixed-size allocator.
We expect this page cache to have a much cheaper allocation cost, at the disadvantage of storing file data in non-linear blocks of memory---observable in the difference between the \deffont{write} and \deffont{update} measurements.
For \deffont{write}, \texttt{tmpfs} outperforms Veracruz due to faster memory allocations and no unnecessary copies, while \deffont{update} requires no memory allocation, and has more comparable performance.

\begin{figure}[t!]
\centering
\vspace*{-4pt}
\begin{tikzpicture}
\begin{axis}[
    ymode=log,
    height=5.5cm,
    xlabel={Dimension of square matrix},
    y tick label style={font=\scriptsize},
    x tick label style={font=\scriptsize},
    ylabel={Exec.~time (ms)},
    y label style={font=\scriptsize},
    x label style={font=\scriptsize},
    legend pos=north west,
    point meta=rawy, 
    nodes near coords,
    nodes near coords align={right},
    nodes near coords style={rotate=90, yshift=0.0cm, xshift=0.05cm, font=\tiny, /pgf/number format/.cd,std=0:3,fixed zerofill,sci zerofill,sci precision=2,precision=2},
    grid=both, 
    grid style=dashed,
    legend style={
        inner xsep=0pt,inner ysep=0pt,nodes={inner xsep=3pt},
        legend columns=1,
        anchor=south east,
        draw=none,
        font=\tiny,
        at= {(current axis.south east)}
        },
    legend cell align=left,
]
\addplot[arm-blue, mark=*] table [x=time, skip first n=4, y expr=\thisrow{computation}/1000]{evaluation/FHE.txt};
\addlegendentry[]{FHE computation};

\addplot[coordinate style/.condition={true}{yshift=-30}, arm-orange, mark=diamond*] table [x=time, skip first n=4, y expr=(\thisrow{enc} + \thisrow{dec})/1000]{evaluation/FHE.txt};
\addlegendentry[]{FHE encryption};

\addplot[arm-grey, mark=square*] table [x=time, skip first n=4, y expr=\thisrow{veracruz}/1000]{evaluation/FHE.txt};
\addlegendentry[]{Veracruz computation};
\end{axis}
\end{tikzpicture}
\vspace{-0.25cm}
\caption{SEAL and Veracruz computation performance}
\label{fig:veracruz.vs.fhe}
\end{figure}
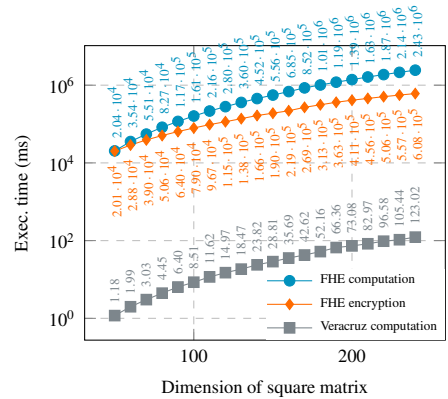

\paragraph{Fully-homomorphic encryption}
An oft-suggested use-case for fully-homomorphic encryption (FHE, hencefoth) is protecting delegated computations.
We briefly compare Veracruz against SEAL~\cite{microsoft:seal:2019}, a leading FHE library, in computing a range of matrix multiplications over square matrices of various dimensions.
Algorithms in both cases are written in \texttt{C}, though floating point arithmetic is replaced by the SEAL multiplication function for use with FHE.
Results are presented in Fig.~\ref{fig:veracruz.vs.fhe}.
Our results demonstrate that overheads for FHE are impractical, even for simple computations.

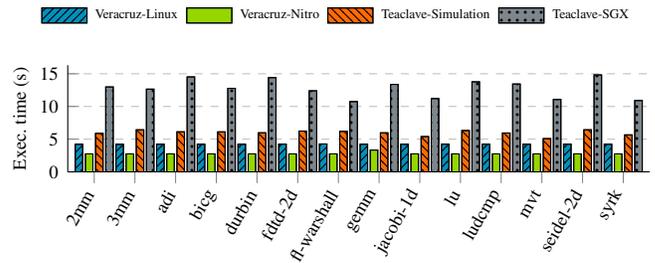
\begin{figure}[t!]
\vspace*{-10pt}
\begin{tikzpicture}

\begin{axis} [
    width=\columnwidth * 1.1, height=3cm, 
    ymajorgrids=true,
    grid style=dashed,
    align=center,
    ybar=1pt, 
    ymin=0,
    axis y line*=none, axis x line*=none,
    bar width = 0.1cm,
    xtick = data,
    symbolic x coords={2mm, 3mm, adi, bicg, durbin, fdtd-2d, fl-warshall, gemm, jacobi-1d, lu, ludcmp, mvt, seidel-2d, syrk}, 
    xmin=2mm,
    xmax=syrk,
    x tick label style={rotate=60,anchor=east,font=\scriptsize},
    y tick label style={font=\scriptsize},
    point meta=rawy, 
    legend style={
        inner xsep=0pt,inner ysep=10pt,nodes={inner xsep=3pt},
        yshift=5pt,
        legend columns=4,
        anchor=south east,
        draw=none,
        font=\tiny,
        align=right,
        },
    legend cell align=left,
    enlarge x limits={rel=0.05},
    xlabel style={yshift=0.0cm},
    ylabel style={yshift=-0.1cm, font=\scriptsize},
    ylabel={Exec. time (s)},
    area legend] 
]

\addplot [black,fill=arm-blue, postaction={pattern = north east lines}] table[skip first n=16, x=prog,y expr=\thisrow{v-linux}/1000] {evaluation/teaclave.txt};
\addlegendentry[]{Veracruz-Linux};
\addplot [black,fill=arm-green, postaction={pattern = none}] table[skip first n=16, x=prog,y expr=\thisrow{v-ec2}/1000] {evaluation/teaclave.txt};
\addlegendentry[]{Veracruz-Nitro};
\addplot [black,fill=arm-orange, postaction={pattern = north west lines}] table[skip first n=16, x=prog,y expr=\thisrow{t-sim}/1000] {evaluation/teaclave.txt};
\addlegendentry[]{Teaclave-Simulation};
\addplot [black,fill=arm-grey, postaction={pattern = dots}] table[skip first n=16, x=prog,y expr=\thisrow{t-sgx}/1000] {evaluation/teaclave.txt};
\addlegendentry[]{Teaclave-SGX};

\end{axis}

\end{tikzpicture}
\spacehack{20pt}
\caption{Execution times of Veracruz and Apache Teaclave}
\label{fig:teaclave.polybench}
\end{figure}

\paragraph{Teaclave}
Apache Teaclave~\cite{apache/incubator-teaclave-sgx-sdk_2020} is a privacy-preserving FaaS infrastructure built on Intel SGX, supporting Python and Wasm with a custom programming model using the Wamr~\cite{wamr} interpreter.
We compare the performance of Teaclave running under Intel SGX with Veracruz as a Linux process, both on \coreiseven, and Veracruz on AWS Nitro enclaves on \ectwo---admittedly an imperfect comparison, due to significant differences in design, isolation technology, Wasm runtime, and hardware between the two.
We run the PolyBench/C suite with its mini dataset---Teaclave's default configuration errors for larger datasets---and measure end-to-end execution time, which includes initialization, provisioning, execution and fetching the results, which we present in Fig.~\ref{fig:teaclave.polybench}.
While Veracruz has better performance than Teaclave when executing Wasm---with Veracruz under AWS Nitro exhibiting a mean $2.11\times$ speed-up compared to Teaclave in simulation mode, and faster still than Teaclave in SGX---the fixed initial overhead of Veracruz, ${\sim}4$s in Linux and ${\sim}2.7$s in AWS Nitro, dominates the overall overhead in either case.

\section{Closing remarks}
\label{sect.closing.remarks}

We have introduced \emph{Veracruz}, a framework for designing and deploying privacy-preserving delegated computations among a group of mutually mistrusting principals, using isolates as a ``neutral ground'' to protect computations from prying or interference.
In addition to supporting a number of hardware-backed Confidential Computing technologies---such as AWS Nitro Enclaves and Arm Confidential Computing Architecture Realms---Veracruz also supports pragmatic ``software isolates'' through \emph{IceCap}. IceCap makes use of the high-assurance seL4 microkernel, on Armv8-A platforms without any other explicit support for Confidential Computing, to provide strong isolation guarantees for virtual machines.

Veracruz, with IceCap, provides a uniform programming and attestation model across emerging and ``legacy'' hardware platforms, easing the deployment of delegated computations.
Both projects are open-source~\cite{icecap-repo, veracruz-repo}, and Veracruz is adopted by the LF's \emph{Confidential Computing Consortium}.

\paragraph{Related work}
Isolates have been used to protect a zoo of computations of interest, e.g.,~ML~\cite{10.5555/3241094.3241143, DBLP:conf/iclr/TramerB19, 10.1007/978-3-319-66402-6_21, 10.1145/3133956.3134095, DBLP:journals/corr/abs-1902-04413} and genomic computations~\cite{10.1007/978-3-030-00305-0_21, feng:presage:2016, 8613959}, and have been used to emulate or speed up cryptographic techniques such as functional encryption~\cite{10.1145/3133956.3134106} and secure multi-party computations~\cite{3f80bb9409ae4c659cc5b1ed26d751f8, 10.1007/978-3-662-53357-4_20, 10.1145/3338466.3358919}.
These can be seen as use-cases, specialized with a particular policy and program, of Veracruz.

\emph{OpenEnclave}~\cite{openenclave} provides a common development platform for SGX Enclaves and TrustZone trusted applications.
Veracruz provides a higher-level of abstraction than OpenEnclave, and includes various support libraries, client code, and attestation protocols to ease the provisioning of programs into an isolate.
Veracruz also supports a wider range of isolates, including both hardware- and software-isolates.

Previous work~\cite{DBLP:conf/trust/KoeberlPRS0Z15} suggested a framework similar to Veracruz, but never implemented it.
\emph{Google Oak}~\cite{google-oak}, \emph{Profian Enarx}~\cite{redhat-enarx}, \emph{Apache Teaclave}~\cite{mesatee}, \emph{Fortanix Confidential Computing Manager}~\cite{fortanix-confidential-computing-manager} and \emph{SCONE}~\cite{DBLP:conf/osdi/ArnautovTGKMPLM16} are similar to Veracruz, though significant differences exist.
Oak's emphasis is information flow control, while Enarx, Fortanix, and SCONE protect the integrity of legacy computations, either requiring recompilation to Wasm, or supporting containerized workloads under SGX, respectively.
Apache Teaclave is the most similar project, discussed in \S\ref{sect.evaluation}, and we perform significantly better.
The proxy attestation service, and our certificate-based attestation protocol, especially suitable for clients on resource-constrained devices, is also unique.

\emph{Protected KVM (pKVM)}~\cite{pkvm, deacon-pkvm} is an attempt to minimize the TCB of KVM, enabling virtualization-based confidential computing on mobile platform, and similar in spirit to IceCap.
pKVM, with an EL2 kernel specifically designed for the task, may have higher performance than IceCap, but will not benefit from the formal verification effort invested in seL4.


\emph{OPERA}~\cite{10.1145/3319535.3354220} places a proxy between client code and the Intel Attestation Service, exposing the same EPID protocol to clients as the web-service exposes.
The Veracruz proxy exposes a potentially different protocol to client code, compared to the native protocol, due to the variety of isolates Veracruz supports.
Intel's \emph{Data Center Attestation Primitives (DCAP)}, also serves similar use-cases, reducing the number of calls to an external attestation service when authenticating attestation tokens, though is limited to use with Intel SGX.

\paragraph{Ongoing and future work}
The proxy attestation service, which currently signs each generated certificate with the same key, could sign certificates for different isolation technologies with different keys, each associated with a different root CA certificate.
With this, a global policy could choose which technology to support based on the selection of root CA certificate embedded in the policy, and if multiple isolation technologies were to be supported, more than one root CA certificate could be embedded.
The proxy attestation server could also maintain multiple Root CA certificates, arranged into a ``decision tree of certificates'', with the server choosing a CA certificate to use when signing the isolate's certificate from the tree, following a path from the root described by characteristics of the isolate technology itself (e.g.,~name of the manufacturer, whether memory encryption is supported, and so on).
Again, the certificate associated with the security profile of the desired isolation technology can be embedded in the policy.

We also aim to bound the intensional and extensional properties of programs provisioned into Veracruz.
Pragmatically, cryptographic operations are perhaps most sensitive to timing attacks, and we aim to provide a limited defense by supplying a constant-time cryptography implementation---using \texttt{mbedtls}~\cite{mbedtls}---via the native module facility discussed in \S\ref{subsect.veracruz.ad.hoc.acceleration}.
Moreover, we aim to explore the use of a statically verified, constant-time virtual machine to gives users the option to statically verify timing properties of their programs---an area of significant recent academic interest---though likely at the cost of limiting their program to constant-time constructs, which is intractable for general-purpose programming.
Using FaCT~\cite{FaCT} Veracruz could provide flexible, verifiably constant-time components such as virtual machines or domain specific functions, while the CT-Wasm~\cite{ct-wasm} extension for Wasm also provides verifiable, constant-time guarantees as a set of secrecy-aware types and bytecode instructions.
CT-Wasm has not yet adopted by the Wasm committee.

We are also continuing work on statically verifying the \emph{Software Fault Isolation} (SFI, henceforth) safety of sandboxed applications.
SFI systems, such as Wasm, add runtime checks to loads, stores, and control flow transfers to ensure sandboxed code cannot escape from its address space region, though bugs in SFI compilers can (and do) incorrectly remove these checks and introduce bugs that let untrusted code escape its sandbox~\cite{bartel2018twenty, cranelift_bug}.
To address this---following other SFI systems~\cite{morrisett2012rocksalt,zhao2011armor,yee2009native}---we have built a static verifier for binary code executed by Veracruz, implemented as an extension of \emph{VeriWasm}~\cite{veriwasm}, an open-source SFI verifier for compiled Wasm code.
To adapt \emph{VeriWasm} to Veracruz, we added support for AArch64, and ported VeriWasm from the \emph{Lucet}~\cite{lucet} toolchain to Wasmtime, as used by Veracruz.
We plan to further extend VeriWasm to check other properties besides software fault isolation, e.g., Spectre~\cite{narayan:2021:swivel} resistance.

Finally, observe that the provisioned program, $\pi$, is either kept classified by its owner, or is declassified to a subset of the other principals in the computation (maybe all).
In the former case, other principals either must either implicitly trust that $\pi$ behaves in a particular way, or establish some other mechanism bounding the behavior of the program, out-of-band of Veracruz.
We aim for a middle ground, allowing a program owner to declassify runtime \emph{properties} of the program, enforced by Veracruz, while retaining secrecy of the program binary (using e.g.,~\cite{supervisionary}).

\inputencoding{utf8}

\end{document}